\documentclass[twocolumn]{aastex61}
\usepackage{color}
\received{}
\revised{}
\accepted{}
\submitjournal{ApJ}

\shorttitle{All-flavor pre-SN neutrino emissions from massive stars}
\shortauthors{Kato et al.}

\begin{document}

\title{Neutrino emissions in all flavors up to the pre-bounce of massive stars and the possibility of their detections}

\correspondingauthor{Chinami Kato}
\email{chinami@heap.phys.waseda.ac.jp}

\author{Chinami Kato}
\affiliation{School of Advanced Science and Engineering, Waseda University, 3-4-1, Okubo, Shinjuku, Tokyo 169-8555, Japan}

\author{Hiroki Nagakura}
\affiliation{TAPIR, Walter Burke Institute for Theoretical Physics, Mailcode 350-17, California Institute of Technology, Pasadena, CA 91125, USA}

\author{Shun Furusawa}
\affiliation{Frankfurt Institute for Advanced Studies, J.W. Goethe University, 60438 Frankfurt am Main, Germany}
\affiliation{Interdisciplinary Theoretical Science Research Group, RIKEN, Wako, Saitama 351-0198, Japan}

\author{Koh takahashi}
\affiliation{Department of Astronomy, The University of Tokyo, Tokyo 113-0033, Japan}
\affiliation{Argelander Institute for Astronomy, Auf dem H$\mathrm{\ddot{u}}$gel 71, 53121 Bonn, Germany}

\author{Hideyuki Umeda}
\affiliation{Department of Astronomy, The University of Tokyo, Tokyo 113-0033, Japan}

\author{Takashi Yoshida}
\affiliation{Department of Astronomy, The University of Tokyo, Tokyo 113-0033, Japan}

\author{Koji Ishidoshiro}
\affiliation{Research Center for Neutrino Science, Tohoku University, Sendai 980-8578, Japan}

\author{Shoichi Yamada}
\affiliation{School of Advanced Science and Engineering, Waseda University, 3-4-1, Okubo, Shinjuku, Tokyo 169-8555, Japan}
\affiliation{Advanced Research Institute for Science and Engineering, Waseda University, 3-4-1, Okubo, Shinjuku, Tokyo 169-8555, Japan}

\begin{abstract}
This paper is a sequel to our previous one \textcolor{black}{\citep{kato15}}, which calculated the luminosities and spectra of electron-type anti-neutrinos ($\bar{\nu}_e$'s) from the progenitors of core-collapse supernovae. Expecting that a capability to detect electron-type neutrinos ($\nu_e$'s) will increase dramatically with the emergence of liquid-argon detectors such as DUNE, we broaden the scope in this study to include all-flavors of neutrinos emitted from the pre-bounce phase. \textcolor{black}{We pick up three progenitor models of an electron capture supernova (ECSN) and iron-core collapse supernovae (FeCCSNe).} We find that the number luminosities reach $\sim10^{57} \mathrm{s^{-1}}$ and $\sim10^{53} \mathrm{s^{-1}}$ at maximum for $\nu_e$ and $\bar{\nu}_e$, respectively. We also estimate the numbers of detection events at terrestrial neutrino detectors including DUNE, taking flavor oscillations into account and assuming the distance to the progenitors to be 200 pc. It is demonstrated that $\bar{\nu}_e$'s from the ECSN-progenitor will be undetected at almost all detectors, whereas we will be able to observe $\gtrsim$15900 $\nu_e$'s at DUNE for the inverted mass hierarchy.
From the FeCCSN-progenitors, the number of $\bar{\nu}_e$ events will be largest for JUNO, 200-900 $\bar{\nu}_e$'s, depending on the mass hierarchy whereas the number of $\nu_e$ events at DUNE is $\gtrsim$2100 for the inverted mass hierarchy. These results imply that the detection of $\bar{\nu}_e$'s is useful to distinguish FeCCSN- from ECSN-progenitors, while $\nu_e$'s will provide us with detailed information on the collapse phase regardless of the type and mass of progenitor.
\end{abstract}


\keywords{stars:evolution --- stars:massive --- supernova:general --- Physical Data and Processes:neutrinos}




\section{Introduction}
Massive stars with $M_{\mathrm{ZAMS}} \gtrsim 8\ M_\odot$ are supposed to be progenitors of core-collapse supernovae (CCSNe), which are violent explosions at the end of their lives. 
The explosion is instigated by the collapse of a central core, which is followed by the formation of a shock wave at core bounce.
If the shock wave runs through the central core and propagates through outer envelopes up to the stellar surface, these envelopes are ejected and a compact remnant is left behind at the center.
How to get the shock wave out of the core has been explored for a long time but has not been settled yet \citep[references therein]{janka12,kotake12}.
One of the current focuses is some features in the structures of progenitors such as the compactness of core and convective activities in the envelopes \citep{muller15,couch15}.

\textcolor{black}{We consider two types of progenitors} that are supposed to produce CCSNe: in the majority case they produce a core mainly composed of irons (Fe-core), which collapses when a certain density or temperature is reached; in the other case, \textcolor{black}{which occupies $\sim$5\% of all CCSNe according to a recent study \citep{doherty17},} the gravitational contraction starts already after a core consisting of oxygens and neons (ONe-core) is formed via carbon burning (C-burning) and grows to a critical mass, $M_{\mathrm{core}} = 1.376\ M_\odot$ \citep{woosley02}. 
The initial stellar mass on the main sequence is the main factor to determine which is obtained in the end: 
stars on the lightest end of massive stars ($\sim$8-10 $M_{\odot}$) will lead to the latter and more massive stars will produce the former \citep{umeda12,jones13}.

In fact, if a star is massive enough ($M_{\mathrm{ZAMS}} \gtrsim 10\ M_\odot$), then the temperature reaches the ignition point of Ne or O at the center, synthesizing iron-group elements through Si-burnings.
Electron captures (ECs) on and/or photo-dissociations of these heavy nuclei trigger the gravitational collapse of Fe-core.
This mode of the evolution to collapse and the ensuing explosion is referred to as ``iron core collapse supernovae (FeCCSNe)''.
For the lighter masses, on the other hand, electrons are more degenerate in the ONe-core and their pressure can support the core even at the vanishing temperature.
The mass of the ONe-core increases through shell C-burnings, however, and if it exceeds the critical value $M_{\mathrm{core}} = 1.376\ M_{\odot}$, at which the central density reaches the threshold for EC on \textcolor{black}{${}^{24}$Mg} ($\log_{10}{\rho_c}/[\mathrm{g\ cm^{-3}}] = 9.88$), then the core begins to contract, losing the pressure support from electrons \citep{koh13}.
This leads in turn to EC on \textcolor{black}{${}^{20}$Ne}, accelerating the contraction and eventually igniting O and Ne.
The O- and Ne-burnings propagate as a deflagration wave, establishing behind it the nuclear statistical equilibrium (NSE).
Neutrinos are then emitted copiously via EC reactions on iron-group elements and free protons, which eventually trigger the collapse of the ONe-core that proceeds on the dynamical time scale.
In this paper we call this mode of collapse and the following CCSN either ``electron capture supernovae (ECSNe)'' or ``ONe-core collapse supernovae (ONeCCSNe)''.
The resultant supernova explosions are supposed to be weaker with an explosion energy of  $\sim10^{50}$ erg than FeCCSNe with $\sim10^{51}$ erg \citep{kitaura06}.
In fact, SN1054, which produced the Crab pulsar, may be one of such ECSNe \citep{nomoto82,tominaga13}.

Neutrinos play an important role in both progenitor evolutions and the supernova explosion itself.
In fact, the neutrino heating mechanism is thought to be the currently most promising scenario to revive a stalled shock and produce a successful explosion.
CCSNe are also one of the most important cosmic neutrino sources from an observational point of view \citep{raffelt12} as corroborated by the observation of neutrinos from SN1987A at terrestrial neutrino detectors such as Kamiokande \citep{hirata1987,arnett1989}.
These neutrinos are mostly emitted in the cooling phase of proto-neutron stars (PNS's), which follows the shock revival and lasts for $\sim$10 secs \citep{sato1987,burrows1988,fischer12}.
Before core collapse, on the other hand, neutrinos dominate photons in the stellar cooling after C-burning.
These neutrinos are called ``pre-supernova (pre-SN) neutrinos''.
As the central temperature and density increase in the progenitor, the number and energy of pre-SN neutrinos also rise \citep{odrzy10}.

\textcolor{black}{Neutrinos are emitted via thermal pair processes and nuclear weak interactions.
Among the former, electron-positron pair annihilations and plasmon decays are important in the late phase of stellar evolution \citep{itoh1996,kato15}.
\cite{odrzy04} were the first to pay attention to the neutrino emissions via the electron-positron pair annihilation and point out that they may be observable during the Si-burning phase if the progenitor is located at a distance $\lesssim$ 1 kpc.
Later they also investigated the energy spectrum of plasmon decay \citep{odrzy07}.
\cite{odrzy09} and \cite{patton15} pointed out that neutrino emissions via nuclear weak processes, such as $\beta^-$ decay, may become dominant just prior to collapse.
\cite{fuller16} discussed the importance of excited states in both parent and daughter nuclei in these processes.}

\cite{kato15} \textcolor{black}{(``Paper I'' hereafter)} took into account realistic stellar evolutions that lead to both the FeCCSN and ONeCCSN.
They showed that these two types of supernova progenitors can be distinguished by the detection (or no detection) of their pre-SN neutrinos.
\cite{yoshida16} investigated more in detail the pre-SN neutrino luminosities and cumulative numbers of detection events as a function of time for FeCCSN-progenitors.
They demonstrated that the pre-SN neutrinos can be used as a useful probe into the Si-burning, which occurs deep inside massive stars, if they are observed on the next-generation detectors such as JUNO and Hyper-Kamiokande.


In the observation of $\bar{\nu}_e$'s, the detectors in operation at present, both water Cherenkov and liquid scintillation types, employ mainly the inverse $\beta$ decay whose cross section dominates those of other reactions such as the elastic scattering on electron.
It is $\nu_e$'s, however, that are produced in the largest quantity as a result of EC.
It is hence nice from the observational point of view that new detectors that have capabilities to detect $\nu_e$ may become available in the near future.
Deep Underground Neutrino Observatory, or DUNE, is a liquid argon detector currently planned to be constructed in 10 years at SURF (Sanford Underground Research Facility) \citep{dune}.
It deploys 4 detectors filled with liquid argon of 10 kt each.
Although the detection of supernova neutrinos emitted after core bounce is one of the main targets of DUNE, it should be noted that its energy threshold will be low enough ($\sim 5$ MeV) to detect $\nu_e$'s in the pre-SN phase.
In this paper we calculate $\nu_e$'s produced via both the thermal and weak processes and discuss their detectability.
Although the Helium and Lead Observatory (HALO) experiment at SNOLAB can also detect $\nu_e$'s with heliums and leads in principle, it is not suitable for the detection of pre-SN neutrinos because of its small volume and high energy threshold \citep{zuber15}.

The neutrino emissions at different phases, i.e., the progenitor phase, pre-/post-bounce phases and PNS-cooling phase, have been investigated separately so far. 
Considering, however, the recent progress in the numerical modeling of CCSNe, in which we observe successful explosions rather commonly, we believe that these phases should be handled consistently, based on successful supernova models. 
This paper is the first step in this direction and we attempt to calculate neutrino emissions from the progenitor stage up to the pre-bounce time, at which the central density becomes $\log_{10}{\rho_c}/[\mathrm{g\ cm^{-3}}] = 13$, consistently and seamlessly.
The subsequent evolutions of the same models will be studied later.


The organization of the paper is as follows: the progenitor models for the ECSN and FeCCSN are briefly described in Section 2;
the calculations of the luminosities and spectra of neutrinos are summarized for individual processes in Section 3; 
the results are presented in Section 4, and finally the summary and discussions are given in Section 5.

\section{models}
In this paper we consider neutrino emissions during both the quasi-static evolutions of progenitors and the hydrodynamical core-collapse. \textcolor{black}{We stop the calculations at the time when the central density reaches $\log_{10}{\rho_c}/[\mathrm{g\ cm^{-3}}] = 13$.} For the former we use the stellar evolution models as described in section 2.1 whereas for the latter we conduct one-dimensional simulations under spherical symmetry, solving radiation-hydrodynamics equations as explained in section 2.2. Note that we need to take into account neutrino transport in the core properly once the density becomes high enough to trap neutrinos.
The two evolutionary phases are connected at the time when the central density becomes $\log_{10}{\rho_c}/[\mathrm{g\ cm^{-3}}] = 10.3$ for FeCCSNe and $10.1$ for ECSNe, respectively.

\subsection{Quasi-static evolutions of progenitors}
We employ three progenitor models with $M_{\mathrm{ZAMS}} = $ 9, 12 and 15 $M_\odot$, which were calculated anew by Takahashi \citep[see][]{koh13,koh16}.
The first one produces an ONe-core that is supposed to explode as ECSN, while the last two models explode as FeCCSNe if they really succeed to.
\textcolor{black}{We employ the 9 $M_\odot$ model instead of the 8.4 $M_\odot$ model adopted in Paper I, since \cite{koh17} have improved the treatment of convective overshooting in the 9 $M_\odot$ model and currently investigating in detail the core collapse and the subsequent explosion of the same model.}
The latter two models with 12 and 15 $M_\odot$ are indeed identical to those employed in \cite{yoshida16} but calculated until the central temperature reaches $10^{10}$ K with hydrodynamics taken into account.

\begin{figure}
\epsscale{1.2}
\plotone{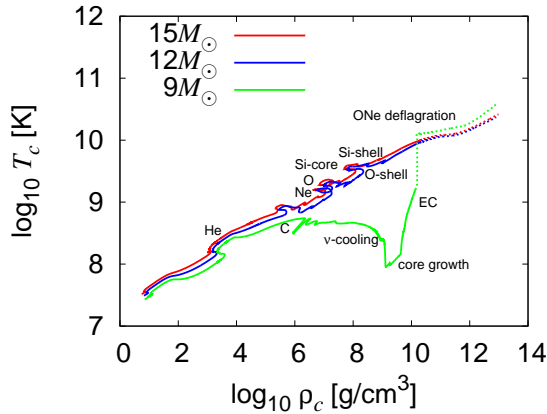}
\caption{The evolutionary paths of the central density and temperature for three progenitors. The red, blue and green curves correspond to the 15, 12 and $9\ M_\odot$ models, respectively. The evolutions in both the progenitor phase (solid lines) and the collapse phase (dotted lines) are presented. \textcolor{black}{The initiation points of some major nuclear-burnings as well as the evolutionary stages defined by \cite{koh13} for the ONe-core progenitors are marked with labels.} \label{fig1}}
\end{figure}

\begin{figure}
\epsscale{1.2}
\plotone{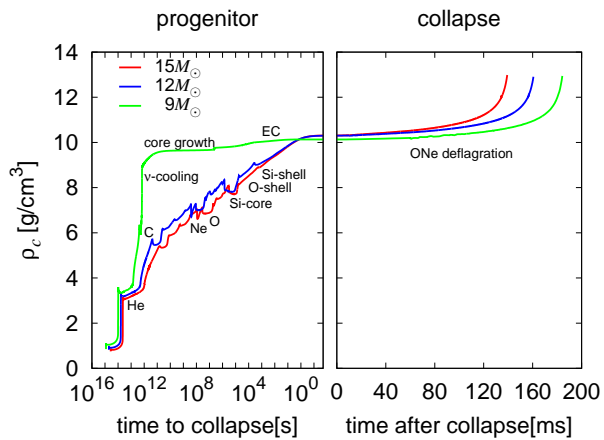}
\caption{The evolutions of the central density as functions of time for the three progenitors. The red, blue and green curves correspond to the 15, 12 and $9\ M_\odot$ models, respectively. The origin of the horizontal axis corresponds to the time, at which the dynamical simulations are started. The initiation points of some major nuclear-burnings \textcolor{black}{as well as the evolutionary stages for the ONe-core progenitors} are marked with labels. \label{fig1.5}}
\end{figure}

\begin{figure*}
\epsscale{0.8}
\plotone{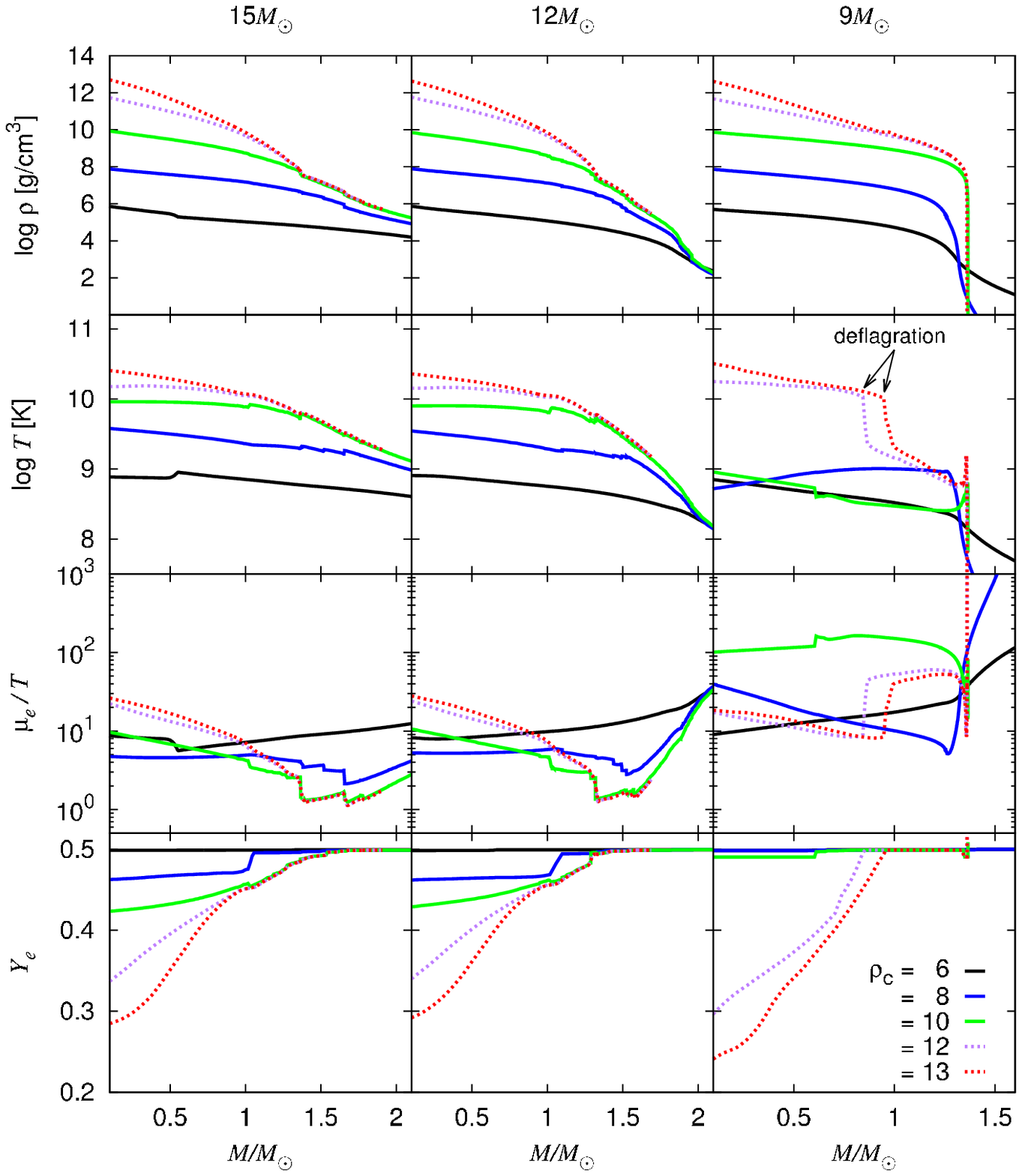}
\caption{The radial profiles of density, temperature, degeneracy, electron fraction at different times. The horizontal axis is the mass coordinate in the solar-mass unit. The left, middle and right columns correspond to the 15, 12 and $9\ M_\odot$ models, respectively. Different colors denote the times, at which the central density becomes $\log_{10}{\rho_c}/[\mathrm{g\ cm^{-3}}]$ = $6$ (black solid), $8$ (blue solid), $10$ (green solid), $12$ (purple dotted) and $13$ (red dotted), respectively. \label{fig2}}
\end{figure*}

Here we summarize the evolutions of these models briefly.
Figure \ref{fig1} shows the evolutions of the central density and temperature of the progenitors.
The solid lines represent the results of the quasi-static stellar-evolutionary calculations, or ``progenitor phase'' , whereas the dashed lines correspond to those of the core-collapse simulations, or ``collapse phase''.
\textcolor{black}{In this figure, we also mark the initiation points of major nuclear-burning stages, which are defined to be the points when the element of relevance is ignited at the center; for the ONe-core case  more detailed evolutionary stages are indicated as well, which are defined in \cite{koh13}.}
We see that the two types of progenitors are not much different up to the end of C-burning ($\log_{10}{\rho}/[\mathrm{g\ cm^{-3}}] \sim 6$).
After that, however, the evolutionary paths are deviated remarkably from each other.
The progenitors with 12 and 15 $M_{\odot}$ proceed further to burn heavier nuclei stably under the supports of not only thermal but also degenerate pressures and their central densities and temperatures increase gradually up to collapse.
In the case of the progenitor with $M_{\rm{ZAMS}} = $ 9 $M_{\odot}$, on the other hand, the Ne-burning does not occur immediately, since the temperature does not become high enough after the C-burning. 
The core is cooled by neutrino emissions and the central temperature is lowered as the ONe-core grows via the shell C-burning and the central density increases. 
When it reaches the critical value ($\log_{10}{\rho}/[\rm{g\ cm^{-3}}] = 9.88$) for the EC on \textcolor{black}{${}^{24}$Mg}, then the core starts to contract with a shorter time scale and the central temperature also begins to rise again. 
The contraction is accelerated considerably when the EC on \textcolor{black}{${}^{20}$Ne} sets in at $\log_{10}{\rho}/[\rm{g\ cm^{-3}}] = 10.3$, accompanied by a rapid rise of the central temperature. 
Finally, Ne and O are ignited at the center almost simultaneously and the flame front starts to propagate outward as a deflagration. 
The temperature increases drastically and the NSE is established soon after the passage of the burning front.

The evolutions of the central density for the three progenitors are shown in Fig.~\ref{fig1.5}. 
The origin of the time coordinate corresponds to the time, at which the hydrodynamical calculations are initiated.
When a new nuclear burning starts, the core expands and the central density is lowered a bit.
It is evident also in this figure that the pre-collapse evolution of the 9 $M_\odot$ progenitor is qualitatively different from the other two.

In Figure \ref{fig2}, the radial profiles of density $\rho$, temperature $T$, electron degeneracy $\mu_e/T$, where $\mu_e$ is the chemical potential of electron, and electron fraction $Y_e$ are plotted.
The horizontal axis is the mass coordinate in the solar mass unit.
Different colors correspond to different times, at which the central densities are $\log_{10}{\rho_c}/[\mathrm{g\ cm^{-3}}]$ = 6, 8, 10, 12 and 13 , respectively.
It is clear from the comparison between the progenitors of FeCCSNe and that of ECSN, the temperature profiles become different qualitatively at $\log_{10}{\rho_c}/[\mathrm{g\ cm^{-3}}] = 8$.
In the case of the 9 $M_\odot$ progenitor, the central part of the core is cooler than the outer part because of the neutrino cooling via plasmon decay.
The degeneracy parameter $\mu_e/T$ is accordingly higher than those in the 12 and 15 $M_{\odot}$ models.
Rather high electron fractions \textcolor{black}{($Y_e \sim 0.498$)} at early times are a noteworthy feature for the ECSN progenitor.
Although EC reactions trigger the core contraction, the change in $Y_e$ is rather minor \textcolor{black}{($\Delta Y_e \sim 0.008$)} in this phase and the main reduction of $Y_e$ occurs only after NSE is established by the O+Ne deflagration.

\subsection{Core collapse}
Once the accelerated gravitational contraction happens after the EC on \textcolor{black}{${}^{20}$Ne} in the core, we have to abandon the quasi-static approximation and need to solve hydrodynamical equations numerically. 
As explained earlier, interactions of neutrinos with matter become non-negligible as the density increases and neutrinos are eventually trapped in the core.
Then we need to take into account the transport of neutrinos appropriately.
We hence employ the 1-dimensional hydrodynamical code with a Boltzmann solver developed by \cite{nagakura14,nagakura16}
to follow the evolution of core collapse.
The hydrodynamics solver is explicit and of second-order accuracy in both space and time, based on the so-called central scheme \citep{kurganov00,nagakura08,nagakura11}; the spherical coordinates are adopted; the Newtonian self-gravity is taken into account.
The Boltzmann solver adopts the discrete-ordinate method, or the $S_N$ scheme \textcolor{black}{\citep{mezzacappa1993,lieben04,sumiyoshi05}}, finite-differencing both space and momentum space; it is semi-implicit in time; special relativity is fully accounted for by utilizing the two different energy grids: Lagrangian-remapped and laboratory-fixed grids.
\textcolor{black}{Although we normally deploy 12-15 energy grid points spaced logarithmically between 1-300 MeV in this sort of simulations, we increase the number to 20, extending at the same time the energy range to lower values 0.1 MeV in this study so that we could obtain better resolution at these low energies.}
See \cite{nagakura14} for more details.

We use Furusawa's EOS \citep{furusawa13}, a multi-nuclear species EOS, which is based on the relativistic mean field theory with the TM1 parameter set employed in H. Shen's EOS, or STOS EOS \citep{shen11}; it takes into account the NSE among \textcolor{black}{$\sim8.7\times 10^5$ nuclides} and nucleons by extending a nuclear mass formula \textcolor{black}{\citep{audi,tachi6}}; electron capture rates for heavy nuclei are also provided by this EOS at high densities (see below).

In the neutrino transport, the following reactions are taken into account in this paper:
\begin{itemize}
\item{neutrino emissions and absorptions: electron captures on nuclei and free nucleons, electron-positron annihilations, nucleon-bremsstrahlungs and their inverse reactions.}
\item{neutrino scatterings: isoenergetic scatterings on free nucleons, coherent scatterings on nuclei and non-isoenergetic scatterings on electrons and positrons.}
\end{itemize}
The reaction rates are based on \cite{bruenn1985} and \cite{mezza1993} except for the EC on heavy nuclei, for which we take the values provided by \cite{fuller1985,oda1994,langanke01} and \cite{langanke03}, which are referred to as FFN, ODA, LMP and LMSH, respectively, and average them over the NSE abundance of nuclei given by the EOS. We also employ the approximation formula (see eq.~(\ref{apro1}) below) when none of the tables provide the rate. 
The luminosity and energy spectrum of $\nu_e$ in the collapse phase are obtained directly from the simulations whereas those for other neutrino species are calculated in the post processes (see the next section). 

We use for the dynamical simulations only the radial profiles of central cores derived from the quasi-static evolutions of the progenitors.
For the Fe-cores of the 12 and 15 $M_\odot$ models, we start the computations from the time when the central density is $\log_{10}{\rho_c}/[\mathrm{g\ cm^{-3}}] = 10.3$.
We first run the Boltzmann solver alone with all quantities other than the neutrino distribution functions being fixed until steady states are reached.
This step is necessary to avoid artificial discontinuities in the neutrino luminosities at the point of the switch to the dynamical calculations.
The 9 $M_\odot$ model needs a special treatment. 
As already mentioned, the gravitational contraction starts in the ONe-core via EC. 
Neons and oxygens are then ignited at the center and the deflagration wave propagates outward in the core, establishing NSE behind. 
Note that NSE already prevails in the Fe-cores. 
In the case of the ONe-core, we hence have to handle this nuclear equilibration process, solving network equations in addition to the hydrodynamics and neutrino transport. 
This has been done recently by \cite{koh17} and we will use their results in this paper.
Since details will be published in their forthcoming paper, we here give important information alone: they modified the 1D radiation-hydrodynamics code developed by \cite{sumiyoshi12} to compute the nuclear reactions with a network of 40 nuclear species;
H. Shen's original EOS is employed instead of Furusawa's extended version;
EC rates are adopted from \cite{juod08}.
The radial profile at the time of $\log_{10}{\rho_c}/[\mathrm{g\ cm^{-3}}] = 10.1$ is used as the initial condition for the simulation.

In all three cases, we terminate the simulations when the central density exceeds $\log_{10}{\rho_c}/[\mathrm{g\ cm^{-3}}] = 13$. 
This is because nuclei become very large thereafter and pastas are supposed to emerge eventually toward core bounce \textcolor{black}{\citep{ravenhall}}; then the EC rates on these nuclei are highly uncertain and treated only crudely in the original radiation-hydrodynamics code.



The dotted lines in Figs.~\ref{fig1} and \ref{fig2} show the evolutions in the collapse phase.
The behavior of the central temperature and density in this phase is not much different between the two types of progenitors.
In the 9 $M_\odot$ progenitor, however, the temperature is high only inside the deflagration front, which is located at the mass coordinate of $\sim1\ M_\odot$ in Fig.~\ref{fig2}.
The NSE condition ($T \gtrsim 5 \times 10^{9}~\mathrm{K}$) is achieved indeed and the degeneracy of electrons is partially lifted there.
It is also evident that EC is drastically enhanced once NSE is established.
Note in passing that the differences in $Y_e$ between the ONe-core and Fe-cores presented here may partially reflect the differences in the EOS and EC rates adopted in these models.


\section{Neutrino emissions}
Neutrinos are emitted via several processes, which are classified into thermal pair emissions and nuclear weak interactions.
In this section, we first describe somewhat in detail the formulae we employ to evaluate the neutrino emissivity for individual processes ($\S\S$ 3.1-3.3).
In order to discuss the possibility of observations at terrestrial detectors, flavor oscillations should be taken into account and will be discussed in Section 3.4.
In the progenitor phase, we simply evaluate the luminosities and spectra of all flavors of neutrinos in post processes, i.e., \textcolor{black}{we extract density-, temperature- and electron fraction profiles from the data obtained in the stellar evolution calculations and core-collapse simulations at appropriate times from $\sim10^6$ s to a few ms before core bounce. Then we calculate the neutrino emissivities for the thermal pair productions and nuclear weak interactions (See Table.\ref{tab1}) pointwise and integrate the results outwards from the center of the star until the number luminosities do not change appreciably $\sim10^{-6}$\%}.

In the collapse phase, on the other hand, we treat $\nu_e$ differently from the other flavors of neutrinos: the luminosity and spectrum of $\nu_e$ are derived directly from the radiation-hydrodynamical simulations, since we have to take into account neutrino transport when the opacity for neutrinos gets high enough to hinder their free escape. 
Note that $\nu_e$'s are responsible for the transfer of the electron-type lepton number and hence the evolution of $Y_e$, and their transport in the core is indispensable for realistic supernova simulations.


Other species of neutrinos in the collapse phase, on the other hand, are treated in post-processes, i.e., we first run simulations neglecting these reactions\footnote{\textcolor{black}{The productions and absorptions of neutrinos via the electron-positron pair annihilations are {\it included} in the simulations of core collapse. The resolution of the energy spectra obtained in the simulations are rather low, however, rough and we re-construct them in the post-process.}}; we then extract the densities, temperatures and electron fractions as well as the distribution function of $\nu_e$ from results of the simulations and plug them into the formulae of emissivities given in the following subsections. 
Note that the distribution function of $\nu_e$ is necessary to take into account the Fermi-blocking in the final state. We ignore the transport of these neutrinos, since they are much less abundant than $\nu_e$.
\textcolor{black}{In fact, we compare the emissivities of $\bar{\nu}_e$ via $\beta^-$ decay inside the opaque part of the core ($\log_{10}\rho/\mathrm{[g/cm^3]} > 11.5$) with those in the whole NSE regions at the time when the central density is $\log_{10}{\rho_c}/[\mathrm{g/cm^3}] = 13$ and find that the former contributes only $\sim$0.001\% to the total neutrino emissivities because of the high-degeneracy of electrons there. The emissions of these neutrinos after matter becomes opaque are hence negligible compared with those before that.}
The neutrino emission processes and their treatments in our calculations are summarized in Table.~\ref{tab1}.

\begin{deluxetable*}{c|c|c|c|ccc}
\tablecaption{Neutrino reactions considered in this paper. \label{tab1}}
\tablewidth{0pt}
\tablehead{
\multicolumn{1}{c|}{} & \multicolumn{2}{c|}{reactions} & \colhead{collapse phase\tablenotemark{a}} & \colhead{colors\tablenotemark{b}}
}
\startdata
thermal processes & pair      & $e^- + e^+ \longrightarrow \nu + \bar{\nu} $  & $\nu_e$:T, others:P & red\\
                & plasmon   & $\gamma^\ast \longrightarrow \nu + \bar{\nu}$ &  - & brown\\ 
\tableline
nuclear processes & EC        & $(Z,A) + e^- \longrightarrow (Z-1,A) + \nu_e$ & T  & black\\
                & $\beta^+$ & $(Z,A) \longrightarrow (Z-1,A) + e^+ + \nu_e$  &  - & purple\\
                & PC        & $(Z,A) + e^+ \longrightarrow (Z+1,A) + \bar{\nu}_e$ &  P & orange\\
                & $\beta^-$ & $(Z,A) \longrightarrow (Z+1,A) + e^- + \bar{\nu}_e$ &  P & green\\
                & free p    & $p + e^- \longrightarrow n + \nu_e$               &  T & blue\\
\enddata
\tablenotetext{a}{The forth column gives the treatment of each process in the collapse phase: T means that the transport is considered whereas P stands for the post-process and - implies that the process is neglected.}
\tablenotetext{b}{The fifth column lists the color-codes used consistently in Figs.~\ref{fig3}-\ref{fig7}.}
\end{deluxetable*}

\subsection{Thermal emissions of neutrino pairs}
For this type of emissions we normally consider four processes: electron-positron annihilation, plasmon decay, bremsstrahlung\footnote{This bremmsstrahlung occurs in association with a collision of an electron with a nucleus via electromagnetic interactions and is different from the bremsstrahlung from nucleon-nucleon collisions via nuclear forces, which becomes important in the post-bounce phase.} and photo process.
They produce all flavors of neutrinos.
The reaction rates of these processes depend mainly on three hydrodynamical variables: density, temperature and electron fraction (or electron chemical potential).

\cite{itoh1996} investigated in detail which process is dominant in which regime.
\textcolor{black}{In Paper I} we found that the electron-positron pair annihilation is always dominant for the FeCCSN-progenitors with 12 and 15 $M_\odot$, while for the ONeCCSN-progenitor with 9 $M_\odot$ the plasmon decay prevails until Ne and O are ignited at the center and the temperature rises quickly, after which the pair annihilation takes over.
In this paper we hence focus on these two processes as in Paper I.
See also \cite{patton15} and \cite{guo16}.

\subsubsection{Electron-positron annihilation}
Neutrino-pair creations through the electron-positron annihilations become important at high temperatures $\gtrsim 10^{9}\ {\rm K}$ simply because the number of photons with high enough energies to produce electron-positron pairs becomes large and, as a result, electron-positron pairs become also abundant at these temperatures.
\textcolor{black}{Detailed derivations of the reaction rate $R$ for the pair annihilation are given in \textcolor{black}{Paper I (Appendix A.1) but with some typographical errors.} 
We give here the correct expression for $R$} \citep{mezza1993,schinder1982}:
\begin{equation}
R=\displaystyle{\frac{8G_F^2}{\left(2\pi\right)^2} \left[\, \beta_1\, I_1 + \beta_2\, I_2 + \beta_3\, I_3\, \right]}. \label{R}
\end{equation}
In this expression, $\beta$'s are the following combinations of the coupling constants:
 $\beta_1=\left(C_V-C_A\right)^2$, $\beta_2=\left(C_V+C_A\right)^2$ and $\beta_3=C_V^2-C_A^2$, 
and $I$'s are the functions of the energies of emitted neutrino $E_{\nu}$ and anti-neutrino $E_{\bar{\nu}}$ and the angle $\theta$ 
between their momenta $\mbox{\boldmath$q$}$ and $\mbox{\boldmath$q$}^\prime$: 
\begin{eqnarray}
&&I_1\left(E_{\nu}, E_{\bar{\nu}}, \cos{\theta} \right)  \nonumber \\
&&\ \ \ = -\frac{2\pi T \, E_{\nu}^2 {E_{\bar{\nu}}}^2 
\left(1-\cos{\theta}\right)^2}{\left[\exp{\left(\frac{E_{\nu}+E_{\bar{\nu}}}{T}\right)} - 1\right]{\Delta_e}^5} \nonumber \\ 
&&\ \ \ \times \left \{ AT^2 \left( \left[G_2\left(y_{\mathrm{max}}\right) - G_2\left(y_{\mathrm{min}}\right) \right] \right. \right. \nonumber \\
&&\ \ \ \ \ \ \ \ \ \ \ \left. \left. + \left[2y_{\mathrm{max}}G_1\left(y_{\mathrm{max}}\right) 
- 2y_{\mathrm{min}}G_1\left(y_{\mathrm{min}}\right)\right] \right. \right. \nonumber \\
&&\ \ \ \ \ \ \ \ \ \ \ \left. \left. + \left[y_{\mathrm{max}}^2G_0\left(y_{\mathrm{max}}\right) 
- y_{\mathrm{min}}^2G_0 \left(y_{\mathrm{min}} \right) \right] \right)  \right. \nonumber \\
&& \ \ \ \ \ \ + BT \left(\left[G_1\left(y_{\mathrm{max}}\right) 
- G_1\left(y_{\mathrm{min}}\right)\right] \right. \nonumber \\
&& \ \ \ \ \ \ \ \ \ \ \  \left. + \left[y_{\mathrm{max}}G_0\left(y_{\mathrm{max}}\right)
- y_{\mathrm{min}}G_0\left(y_{\mathrm{min}} \right) \right] \right) \nonumber \\
&& \ \ \ \ \ \ \left. +C\left[G_0\left(y_{\mathrm{max}}\right)
-G_0\left(y_{\mathrm{min}}\right)\right] \right \}, \\
&&I_2=I_1\left(E_{\bar{\nu}},E_{\nu},\cos{\theta}\right), \\
&&I_3=-\frac{2\pi  T \, m_e^2\, E_{\nu}{E_{\bar{\nu}}}
\left(1-\cos{\theta}\right)}{\left[\exp{\left(\frac{E_{\nu}+E_{\bar{\nu}}}{T}\right)} - 1\right] \Delta_e} \nonumber \\
&&\ \ \ \times \left[G_0\left(y_{\mathrm{max}}\right)-G_0\left(y_{\mathrm{min}}\right)\right], 
\end{eqnarray}
with 
\begin{eqnarray}
&& {\Delta_e}^2 \equiv  {E_{\bar{\nu}}}^2+{E_{\nu}}^2+2E_{\nu}E_{\bar{\nu}}\cos{\theta}, \\
&& A  =  {E_{\bar{\nu}}}^2+{E_{\nu}}^2-E_{\nu}E_{\bar{\nu}}\left(3+\cos{\theta}\right),  \\
&& B  =  \left[-2{E_{\nu}}^2+{E_{\bar{\nu}}}^2\left(1+3\cos{\theta}\right) \nonumber \right.\\
&&\ \ \ \ \ \left. +E_{\nu}E_{\bar{\nu}}\left(3-\cos{\theta}\right) \right]E_\nu, \\
&& C  =  \displaystyle{\left[\left(E_{\nu}+E_{\bar{\nu}}\cos{\theta}\right)^2
-\frac{1}{2}{E_{\bar{\nu}}}^2\left(1-\cos^2{\theta}\right)\right.} \nonumber\\
&&\ \ \ \ \ \displaystyle{\left. - \frac{1}{2}\left(\frac{m_e\Delta_e}{E_{\nu}}\right)^2\frac{1+\cos{\theta}}{1-\cos{\theta}} \right]{E_\nu}^2},
\end{eqnarray}
and $\eta^{\prime}=\left(\mu_e+E_{\nu}+E_{\bar{\nu}}\right)/T$, $\eta=\mu_e/T,y_{\mathrm{max}}=E_{\mathrm{max}}/T$, $y_{\mathrm{min}}=E_{\mathrm{min}}/T$ and 
$G_n\left(y\right)\equiv F_n\left(\eta^{\prime}-y\right)-F_n\left(\eta-y\right)$, in which the Fermi integral $F_n(z)$ is defined as 
\begin{equation}
F_n\left(z \right)=\int_{0}^{\infty}\frac{x^n}{e^{x-z}+1}dx.
\end{equation} 

The differential number emissivity for neutrino or anti-neutrino, \textcolor{black}{$dQ_N^{\nu_i}/dE_\nu$, in the progenitor phase} is simply given as an integral of $R$ over the momentum of the partner. \textcolor{black}{We employ the spherical coordinates in the momentum space (See Paper I for details). The energy integral was then truncated at some maximum values, which are determined empirically from the temperature.
Our results are in good agreement with those derived with the Monte Carlo method in \cite{yoshida16} within errors of 4.5\%.}

In the collapse phase as the matter density increases and the neutrino energy rises, interactions between matter and neutrinos become no longer ignored.
Electron-type neutrinos, the most abundant species, are eventually trapped in the core at $\log_{10}{\rho}/[\mathrm{g\ cm^{-3}}] \gtrsim 11$ and become degenerate.
Then the pair creation of $\nu_e$ and $\bar{\nu}_e$ is suppressed by the Fermi-blocking in the final state.
Considering the inverse process, we should hence modify the differential emissivity of $\bar{\nu}_e$ in this phase as
\begin{eqnarray}
&&\frac{dQ^{\bar{\nu}_e}_N}{dE_{\bar{\nu}_e}d\cos{\theta_{\bar{\nu}_e}}d\phi_{\bar{\nu}_e}} \nonumber \\
&& =\frac{E_{\bar{\nu}_e}}{2\left(2\pi\right)^3}
\int \frac{d^3 q_{\nu_e}}{\left(2\pi\right)^32E_{\nu_e}} \nonumber \\
&&\ \ \ \times\left[R^p\left(E_{\nu_e},E_{\bar{\nu}_e},\cos{\theta}\right) \nonumber 
\left[1-f_{\nu_e}\left(E_{\nu_e},\theta_{\nu_e}\right)\right]
\left[1-f_{\bar{\nu}_e}\left(E_{\bar{\nu}_e},\theta_{\bar{\nu}_e}\right)\right] \right. \nonumber\\
&&\ \ \ \ \left. - R^a\left(E_{\nu_e},E_{\bar{\nu}_e},\cos{\theta}\right)
f_{\nu_e}\left(E_{\nu_e},\theta_{\nu_e}\right)
f_{\bar{\nu}_e}\left(E_{\bar{\nu}_e},\theta_{\bar{\nu}_e}\right)\right], \label{nu1}
\end{eqnarray}
where $f_{\nu_e}$ and $f_{\bar{\nu}_e}$ are the distribution functions of $\nu_e$ and $\bar{\nu}_e$, respectively.
The direction of neutrino momentum is specified with the zenith and azimuth angles ($\theta_{\nu}$,$\phi_{\nu}$) with respective to the local radial direction.
The first term in the integrand on the right hand side is the production rate whereas the second term represents the absorption rate for the inverse reaction: $R_p$ is given by eq.~(\ref{R}) while $R_a$ is obtained from $R_p$ via the detailed balance condition: $R_a = R_p\mathrm{exp}((E_\nu+E_{\bar{\nu}})/T)$.
We make an approximation $1-f_{\bar{\nu}_e}\left(E_{\bar{\nu}_e},\theta_{\bar{\nu}_e}\right) \sim 1$, which is well justified as $f_{\bar{\nu}_e}\left(E_{\bar{\nu}_e},\theta_{\bar{\nu}_e}\right)$ is small in the collapse phase.

Moreover, we have to take into account matter motions in the collapse phase and distinguish the global inertial frame, or the observer's frame, from the local fluid-rest frame, since the emissivities we have presented so far are all valid in the latter frame.
The emissivities in the observer's frame is obtained by the following transformation:
\begin{eqnarray}
\frac{dQ^{\bar{\nu}_e}_N}{dE_{\bar{\nu}_e}^{\mathrm{lab}}d\cos{\theta_{\bar{\nu}_e}^{\mathrm{lab}}}d\phi_{\bar{\nu}_e}^{\mathrm{lab}}}  
= J\frac{dQ^{\bar{\nu}_e}_N}{dE_{\bar{\nu}_e}^{\mathrm{fr}}d\cos{\theta_{\bar{\nu}_e}^{\mathrm{fr}}}d\phi_{\bar{\nu}_e}^{\mathrm{fr}}}, \label{nu2}
\end{eqnarray}
where the superscipts ``lab'' and ``fr'' stand for quantities in the laboratory and fluid-rest frames, respectively, and $J$ is the Jaccobian:
\begin{eqnarray}
J = \frac{\partial \left(E_{\bar{\nu}_e}^{\mathrm{fr}}, \cos{\theta_{\bar{\nu}_e}^{\mathrm{fr}}}, \phi_{\bar{\nu}_e}^{\mathrm{fr}}\right)}{\partial \left(E_{\bar{\nu}_e}^{\mathrm{lab}}, \cos{\theta_{\bar{\nu}_e}^{\mathrm{lab}}}, \phi_{\bar{\nu}_e}^{\mathrm{lab}} \right)},
\end{eqnarray}
for the following transformations:
\begin{eqnarray}
E^{\rm{fr}} &=& E^{\rm{lab}}\gamma \left(1 - \vec{n}^{\rm{lab}}\cdot \vec{v}\right), \\
\vec{n}^{\rm{fr}} &=&                      
 \frac{1}{\gamma \left(1-\vec{n}^{\rm{lab}}\right)} \left[\vec{n}^{\rm{lab}} + \left(-\gamma + \frac{\gamma-1}{v^2}\vec{v}\cdot\vec{n}^{\rm{lab}}\right) \right],
\end{eqnarray}
with $\vec{n}^{\rm{lab}} = (\sin{\theta_\nu}\cos{\phi_\nu}, \sin{\theta_\nu}\sin{\phi_\nu}, \cos{\theta_\nu})$ being the propagation direction of neutrino.

\subsubsection{plasmon decay}
\textcolor{black}{The plasmon decay is one of the main cooling processes in massive stars after C-burning.}
As we explained already, it is the dominant neutrino-emitting reaction in the ONe-core until NSE is established. 

\textcolor{black}{Although the reaction rates for the plasmon decay were given in Paper I (Appendix A.2), we give the expression of $R$ for completeness:}
\begin{eqnarray}
&&R=\left(\frac{G_F}{\sqrt{2}}\right)^2\frac{16{C_V}^2}{e^2}\frac{2{E_\nu}^2{E_{\bar{\nu}}}^2
\left(1-\cos{\theta}\right)}{\left[1-\exp{\left(\frac{E_\nu+E_{\bar{\nu}}}{T}\right)}\right]} \nonumber \\
&&\times \left \{\frac{3{\omega_p}^2}{{\Delta_e}^2} \delta\left(f_L\left(E_\nu, E_{\bar{\nu}}, \cos{\theta}\right)\right) \right. \nonumber \\
&&\left. \ \ \ \ \ \times \left[\frac{E_\nu+E_{\bar{\nu}}}{2\Delta_e}\ln{\frac{E_\nu+E_{\bar{\nu}}-\Delta_e}{E_\nu+E_{\bar{\nu}}+\Delta_e}} +1\right]  \right. \nonumber \\
&&\left. \ \ \ \ \ \times \left[-2\cos{\theta}\left(E_\nu+E_{\bar{\nu}}\right)^2-2E_\nu E_{\bar{\nu}}\sin^2{\theta} \right. \right.  \nonumber \\
&&\left. \left. \ \ \ \ \ \ \ \ \ \ \ \ +\frac{2\left(E_\nu+E_{\bar{\nu}}\right)^2}{{\Delta_e}^2}
\left(E_\nu+E_{\bar{\nu}}\cos{\theta}\right)\left(E_{\bar{\nu}}+E_\nu\cos{\theta}\right)\right]  \right. \nonumber \\
&&\left. \ \ -\frac{3{\omega_p}^2\left(E_\nu+E_{\bar{\nu}}\right)^2}{{\Delta_e}^2} 
\delta\left(f_T\left(E_\nu, E_{\bar{\nu}}, \cos{\theta}\right)\right) \right. \nonumber \\
&&\left. \ \ \ \ \ \times \left[1+\frac{E_\nu E_{\bar{\nu}}\left(1-\cos{\theta}\right)}{\left(E_\nu+E_{\bar{\nu}}\right)\Delta_e}
\ln{\frac{E_\nu+E_{\bar{\nu}}-\Delta_e}{E_\nu+E_{\bar{\nu}}+\Delta_e}}\right] \right. \nonumber \\
&&\ \ \ \ \ \left.  \times\left[1-\frac{\left(E_\nu\cos{\theta}+E_{\bar{\nu}}\right)\left(E_{\bar{\nu}}\cos{\theta}+E_\nu\right)}{{\Delta_e}^2}\right]  \right \} \label{pla}
\end{eqnarray}
with the following definitions of $f_L\left(E_\nu, E_{\bar{\nu}}, \cos{\theta}\right)$ and $f_T\left(E_\nu, E_{\bar{\nu}}, \cos{\theta}\right)$:
\begin{eqnarray}
&&f_L\left(E_\nu, E_{\bar{\nu}}, \cos{\theta}\right)=2E_\nu E_{\bar{\nu}}\left(1-\cos{\theta}\right) \nonumber \\
&&+\frac{6{\omega_p}^2E_\nu E_{\bar{\nu}}\left(1-\cos{\theta}\right)}{{\Delta_e}^2}\left[\frac{E_\nu+E_{\bar{\nu}}}{2\Delta_\nu}\ln{\frac{E_\nu+E_{\bar{\nu}}-\Delta_e}{E_\nu+E_{\bar{\nu}}+\Delta_e}}+1 \right], \nonumber \\
\\
&&f_T\left(E_\nu, E_{\bar{\nu}}, \cos{\theta}\right) \nonumber \\
&&=2E_\nu E_{\bar{\nu}}\left(1-\cos{\theta}\right) \nonumber \\
&&-\frac{3{\omega_p}^2\left(E_\nu+E_{\bar{\nu}}\right)^2}{2{\Delta_e}^2}\left[1+\frac{E_\nu E_{\bar{\nu}}\left(1-\cos{\theta}\right)}{\left(E_\nu+E_{\bar{\nu}}\right)\Delta_e}
\ln{\frac{E_\nu+E_{\bar{\nu}}-\Delta_e}{E_\nu+E_{\bar{\nu}}+\Delta_e}}\right]. \nonumber \\
\end{eqnarray}
Note that the dispersion relations of the longitudinal and transverse modes are obtained from $f_L = 0$ and $f_T=0$, respectively.

The differential and total emissivities are defined in the same way as for the pair annihilation.
\textcolor{black}{The Dirac deltas in the integrand eq.(\ref{pla}) used in the angular integral.}
Note that neutrinos emitted by plasmon decay have low energies $E_{\nu} \sim 0.5$ MeV and their contribution to the observable luminosity is minor even in the pre-collapse phase (See Paper I for details).
This is even more so in the collapse phase.
We hence estimate only the maximum luminosities for the plasmon decay in the collapse phase, ignoring the Fermi-blocking in the final state.

\subsection{Nuclear weak interactions}
This is the new stuff in this paper, which was ignored in \textcolor{black}{Paper I}.
In the late evolutionary phase of progenitors and during the collapse phase, nuclear weak interactions can no longer be neglected.
In particular, once opened, EC's by heavy nuclei are the dominant reactions.
They play an important role in the hydrodynamics of core-collapse as explained earlier.
Although $\beta^+$ decays of heavy nuclei also emit $\nu_e$'s, they are certainly is sub-dominant.
Electron-type antineutrinos are emitted either by positron captures (PC) or $\beta^-$ decays.
Although they never affect the core dynamics up to bounce, they are important from the observational point of view, since water Cherenkov detectors mainly observe them. 
Moreover, \cite{patton15} pointed out that there may be a period, in which the $\beta^-$ decay dominates the pair annihilation in the production of $\bar{\nu}_e$'s.

\begin{figure}
\epsscale{2.0}
\plottwo{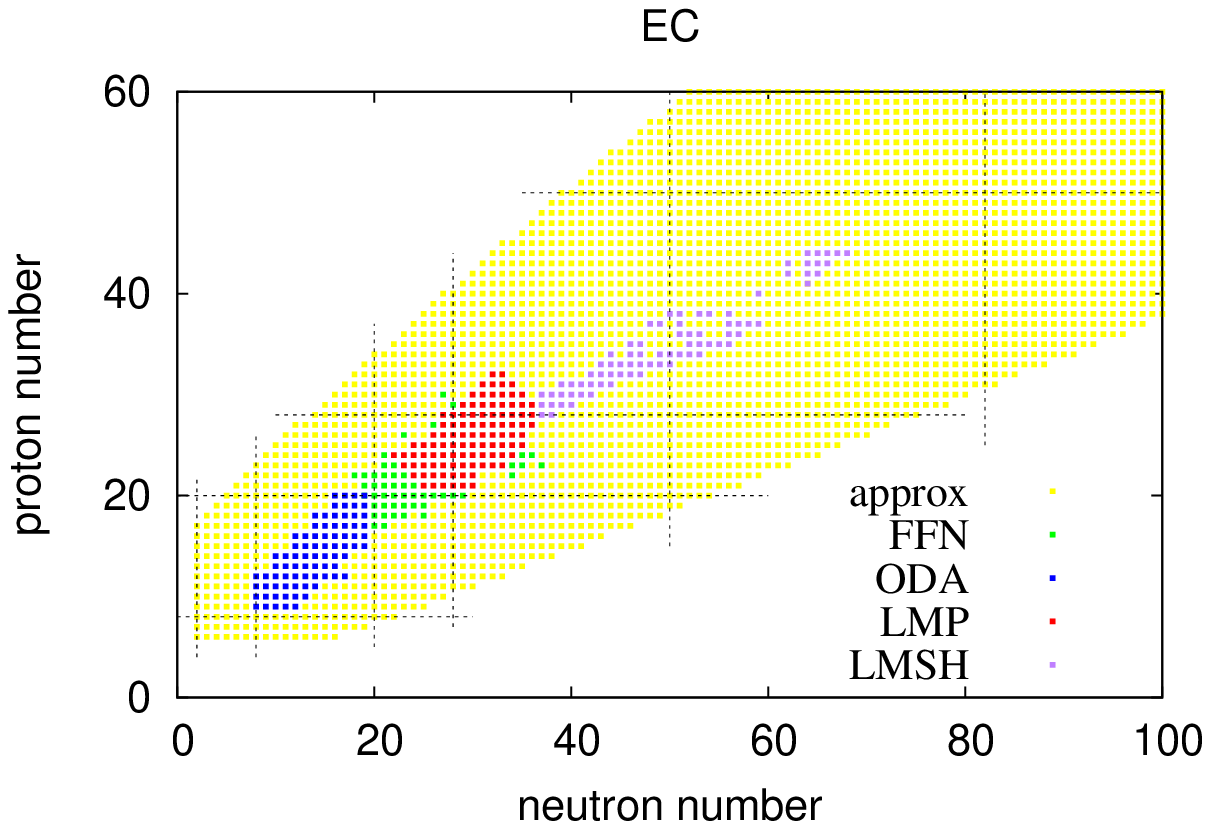}{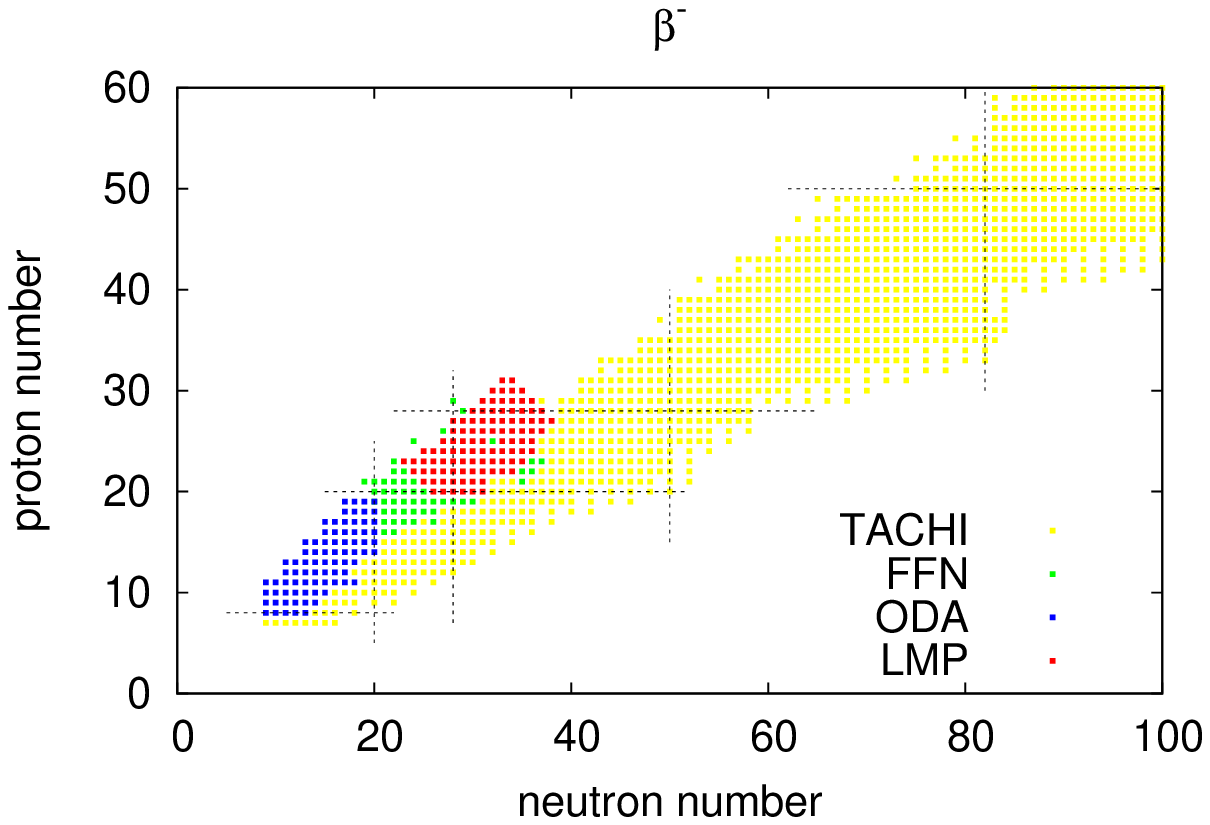}
\caption{Nuclear charts indicating in different colors the nuclear species with the reaction rates for EC (left) and $\beta^-$ decay (right) given in the LMSH (black, \cite{langanke03}), LMP (red, \cite{langanke01}), ODA (blue, \cite{oda1994}), FFN (green, \cite{fuller1985}), TACHI  tables (yellow, \cite{tachi}; \cite{tachi2}; \cite{tachi3}; \cite{tachi4}; \cite{tachi5}; \cite{tachi6}) as well as by the approximate formula (yellow, eq.(\ref{apro1})).  \label{fig3.5}}
\end{figure}

In this paper we hence take into account the following reactions:
\begin{enumerate}
\item{electron capture (EC)}
\begin{equation}
(Z,A) + e^- \longrightarrow (Z-1,A) + \nu_e
\end{equation}

\item{$\beta^+$ decay}
\begin{equation}
(Z,A) \longrightarrow (Z-1,A) + e^+ + \nu_e
\end{equation}

\item{positron capture (PC)}
\begin{equation}
(Z,A) + e^+ \longrightarrow (Z+1,A) + \bar{\nu}_e
\end{equation}

\item{$\beta^-$ decay}
\begin{equation}
(Z,A) \longrightarrow (Z+1,A) + e^- + \bar{\nu}_e.
\end{equation}
\end{enumerate}
In the above expressions, $Z$ and $A$ are the atomic and mass numbers of nuclei, respectively.
\textcolor{black}{We consider in this paper 17502 nuclei (6 $<Z<$ 160, 2 $<N<$ 320) for EC and 3928 nuclei (7 $<Z<$ 117, 9 $<N<$ 200) for $\beta^-$ decay (See also Fig.~\ref{fig3.5}).}

For the calculations of the luminosities and energy spectra of neutrinos we use FFN, ODA, LMP and LMSH tables whenever available.
They normally give us the total reaction rates and average neutrino energies.
If more than one tables are available for the same nucleus, we adopt one of them in the following order: LMSH $>$ LMP $>$ ODA $>$ FFN.
Note that the LMSH table includes data only on the $\nu_e$ emission via EC.
If no information is available from these tables, which actually happens particularly when very heavy and/or neutron-rich nuclei ($A , N $) become populated at late times in the collapse phase, we employ the approximation formulae for $Q_{N,\mathrm{EC}}$ and $Q_{E,\mathrm{EC}}$ \citep{fuller1985,langanke03,sullivan16}:
\begin{eqnarray}
Q_{N,\mathrm{EC}}^{\nu_e} &=& \sum_{i}\frac{X_i\rho}{m_pA_i} \frac{\ln{2}\cdot B}{K}\left(\frac{T}{m_ec^2}\right)^5 \nonumber \\
&&\times\left[F_4\left(\eta\right)-2\chi F_3\left(\eta\right)+\chi^2F_2\left(\eta\right)\right], \label{apro1}
\end{eqnarray}
\begin{eqnarray}
Q_{E,\mathrm{EC}}^{\nu_e} &=& \sum_{i}\frac{X_i\rho}{m_pA_i}\frac{\ln{2}\cdot B}{K}\left(\frac{T}{m_ec^2}\right)^6 \nonumber \\
&&\times\left[F_5\left(\eta\right)-2\chi F_4\left(\eta\right)+\chi^2F_3\left(\eta\right)\right], \label{apro2}
\end{eqnarray}
where $K = 6146$ s, $\chi = (Q - \Delta E)/T$, $\eta = (\mu_e + Q - \Delta E)/T$;
$X_i$ and $A_i$ are the mass fraction and mass number of nuclear species $i$, respectively;
the representative values of the matrix element and the energy level difference between the parent and daughter nuclei are set to $B = 4.6$ and $\Delta E = E_f - E_i = 2.5$ MeV, respectively, following \cite{langanke03}.
For $\beta^-$ decay in the absence of data, we consult another table compiled by Tachibana \citep{tachi,tachi2,tachi3,tachi4,tachi5,tachi6}.
Note that the data in this table were theoretically calculated for the terrestrial environment and hence do not take into account the Fermi-blocking of electrons in the final state. \textcolor{black}{We hence re-incorporated them, albeit crudely, them in the reaction rates as a suppression factor $1-f_e(\langle E_e\rangle)$ based on the average electron energy $\langle E_e\rangle$, which is given in the Tachibana table.}
\textcolor{black}{In Fig.~\ref{fig3.5} we summarize which tables or the approximate formula is used in which region in the nuclear chart.}

The energy spectrum is reconstructed for each reaction by using the effective $q$-value method \citep{langanke01b,kunugise07,patton15}:
\begin{eqnarray}
\frac{dQ_{N,k}^{\nu_j}}{dE_{\nu_j}} &=& N_k\frac{{E_{\nu_j}}^2\left(E_{\nu_j}-q\right)^2}{1+\exp{\left(\frac{E_{\nu_j}-q-\mu_e}{T}\right)}} \nonumber \\
&&\times \Theta \left(E_{\nu_j}-q-m_e\right),
\end{eqnarray}
for $k$ = EC, PC and
\begin{eqnarray}
\frac{dQ_{N,k}^{\nu_j}}{dE_{\nu_j}}  &=& N_k\frac{{E_{\nu_j}}^2\left(q-E_{\nu_j}\right)^2}{1+\exp{\left(\frac{E_{\nu_j}-q+\mu_e}{T}\right)}} \nonumber \\
&&\times \Theta\left(q-m_e-E_{\nu_j}\right),  
\end{eqnarray}
for $k$ = $\beta^-,\beta^+$, where $\nu_j=\nu_e$ or $\bar{\nu}_e$ and the normalization factor $N_k$ is determined by the following relation:
\begin{equation}
Q_{N,k} = \int \frac{dQ_{N,k}^{\nu_j}}{dE_{\nu_j}} dE_{\nu_j}.
\end{equation} 
The effective $q$-value is actually given from the average energy $\langle E_{\nu_e} \rangle$ as follows:
\begin{eqnarray}
\frac{Q_{E,\mathrm{EC}}^{\nu_e} + Q_{E,\beta^+}^{\nu_e}}{\lambda_{\mathrm{EC}} + \lambda_{\beta^+}} 
&=& \langle E_{\nu_e} \rangle \nonumber \\
&=& \displaystyle{\frac{\int  E_{\nu_e} \frac{dQ_{N}^{\nu_e}}{dE_{\nu_e}} \left(E_{\nu_e}\right) dE_{\nu_e} }
{ \int \frac{dQ_{N}^{\nu_e}}{dE_{\nu_e}} \left(E_{\nu_e}\right) dE_{\nu_e}}},
\end{eqnarray}
where the following notation is used:
\begin{eqnarray}
\frac{dQ_{N}^{\nu_e}}{dE_{\nu_e}} \left(E_{\nu_e}\right) = 
\frac{dQ_{N,\mathrm{EC}}^{\nu_e}}{dE_{\nu_e}}
+ \frac{dQ_{N,\beta^+}^{\nu_e}}{dE_{\nu_e}}.
\end{eqnarray}
For $\bar{\nu}_e$, we replace the subscripts of EC and $\beta^+$ with PC and $\beta^-$.


\subsection{Electron capture on free proton}
In the collapse phase, although they are not abundant, EC's on free protons:
\begin{equation}
p + e^- \longrightarrow n + \nu_e,
\end{equation}
cannot be ignored, since the cross section is larger than those of EC's on heavy nuclei.

The reaction rate is given by \cite{bruenn1985} as
\begin{eqnarray}
&&\frac{dQ_{N,\mathrm{p}}^{\nu_e}}{dE_{\nu_e}} = \frac{{G_F}^2}{\pi}\eta_{\mathrm{pn}}\left({g_V}^2+3{g_A}^2\right)\left(E_{\nu_e}+Q\right)^2 \nonumber \\
&&\ \ \ \ \ \ \ \ \times \sqrt{1-\frac{m_e^2}{\left(E_{\nu_e}+Q\right)^2}}f_e\left(E_{\nu_e}+Q\right), \ \ \ \ 
\end{eqnarray}
where the mass difference between neutron and proton is given as $Q=m_n-m_p$, and the form factors for the vector and axial vector currents are given as $g_V=1$ and $g_A=1.23$, respectively; $\eta_{\mathrm{pn}}$ is defined as 
\begin{eqnarray}
\eta_{\mathrm{pn}} &\equiv& \int \frac{2d^3p}{\left(2\pi\right)^3}\tilde{F}_p\left(\tilde{E}\right)\left[1-\tilde{F}_n\left(\tilde{E}\right)\right] \nonumber \\
 &=&\displaystyle{\frac{n_n-n_p}{\exp{\left(\frac{\mu_n^0-\mu_p^0}{T}\right)}-1}.}
\end{eqnarray}
In the above expression, the Fermi-Dirac distributions are denoted by $\tilde{F}_i(\tilde{E}) = 1/[1+\exp{(\tilde{E}-\mu_i)/T}]$ ($i=p,n$), and the number densities and chemical potentials not including the rest-mass energies of proton and neutron are written as $n_n, n_p$ and $\mu_p^0, \mu_n^0$, respectively; the non-relativistic expression $\tilde{E}\sim p_i^2/2m$ is employed for the kinetic energies of nucleons.

\textcolor{black}{In our calculations, the PC and $\beta^-$ decay on neutrons were ignored because they make very little contributions. This is simply because the free neutron is scarce. In addition, the $\beta^-$ decay of free neutron is severely suppressed by the Fermi-blocking of electrons in the final state. Note also that the energy of neutrinos emitted by free neutrons are lower than those by nuclei.}

\subsection{Neutrino oscillations}
The electron neutrinos and anti-neutrinos may convert to $\nu_x$'s and $\bar{\nu}_x$'s, respectively, and vice versa during propagation as a result of flavor oscillations.
We take into account only the vacuum oscillations and MSW effect and ignore the collective oscillations, which will probably not occur in the pre-bounce phase. 
The so-called survival probabilities of $\nu_e$ and $\bar{\nu}_e$ denoted by $p$ and $p^\prime$, respectively, are given in the adiabatic limit as
\begin{equation}
p = \left\{\begin{array}{l}
\sin^2 \theta_{13} = 0.0234\ \ \ \ \ \ \ \ \ \ \ \ \ \mbox{for normal hierarchy,} \\
\sin^2\theta_{12} \cos^2 \theta_{13} = 0.300\ \ \ \mbox{for inverted hierarchy,}
\end{array} \right.
\end{equation}
\begin{equation}
p^\prime = \left\{\begin{array}{l}
\cos^2\theta_{12} \cos^2 \theta_{13} = 0.676\quad \mbox{for normal hierarchy,}\\
\sin^2 \theta_{13} = 0.0234\quad\quad\quad\quad\   \mbox{for inverted hierarchy,}
\end{array} \right.
\end{equation}
with $\cos^2{\theta_{12}}=0.692, \cos^2{\theta_{13}}=0.977$ \citep{pdg14}.
\textcolor{black}{The definition of the mixing angles is common and given in Paper I.}

Since stars are not homogeneous, we need to calculate the number and energy emissivities per volume and time, $Q_{N}^{\nu}$ and $Q_{E}^{\nu}$, as well as the spectra,
$dQ_{N}^{\nu}/dE_{\nu}$, as a function of radius and integrate them over the star to obtain the number and energy luminosities, $L_{N}^{\nu}$ and $L_{E}^{\nu}$, together with the observed spectra, $dL_{N}^{\nu}/dE_{\nu}$ \textcolor{black}{for all flavors of neutrinos in the progenitor phase and for neutrinos other than $\nu_e$ in the collapse phase.
We take the stellar radius as the upper limit of the integrals in principle although the integration started from the center is terminated at some radius when the value does not change appreciably any longer. For the nuclear weak processes, we take the upper limit as the radius of NSE region. 
We evaluate above quantities at different times so that their time evolutions could be obtained.}

As for the pair processes \textcolor{black}{in the collapse phase}, we need actually to conduct two more integrals concerning the zenith and azimuth angles (see eq.~(\ref{nu1})). In so doing, we distinguish the observer's frame from the local fluid-rest frame in the collapse phase (see eq.~(\ref{nu2})).
Then the differential and total number luminosities are given as follows:
\begin{eqnarray}
\frac{dL_N^{\bar{\nu}_e}}{dE_{\bar{\nu}_e}^{\mathrm{lab}}} 
&=&\int \left. \frac{dQ^{\bar{\nu}_e}_N}{dE_{\bar{\nu}_e}^{\mathrm{lab}}d\cos{\theta_{\bar{\nu}_e}^{\mathrm{lab}}}d\phi_{\bar{\nu}_e}^{\mathrm{lab}}}\right|_{\theta_{\bar{\nu}_e}^{\mathrm{lab}} = \theta_s, \phi_{\bar{\nu}_e}^{\mathrm{lab}} = 180^{\circ}} dV \nonumber \\
&=& \int \left.  \frac{dQ^{\bar{\nu}_e}_N}{dE_{\bar{\nu}_e}^{\mathrm{lab}}d\cos{\theta_{\bar{\nu}_e}^{\mathrm{lab}}}d\phi_{\bar{\nu}_e}^{\mathrm{lab}}}\right|_{\theta_{\bar{\nu}_e}^{\mathrm{lab}} = \theta_s, \phi_{\bar{\nu}_e}^{\mathrm{lab}} = 180^{\circ}} \nonumber \\
&& \ \ \ \ \ \ \ \ \ \ \times 2\pi r^2 dr d\cos{\theta_s}, \label{eq:spe_collapse} \ \ \ \\
L_N^{\bar{\nu}_e} &=& 
\int \frac{dL_N^{\bar{\nu}_e}}{dE_{\bar{\nu}_e}^{\mathrm{lab}}} dE_{\bar{\nu}_e}^{\mathrm{lab}}.
\label{eq:lum_collapse}
\end{eqnarray}
In writing these expressions, we assume that the observer is located at infinity on the positive z-axis. 
Note that we employ these formulae only for the electron-positron annihilation, since it is dominant over the plasmon decay in the collapse phase. 
As for the latter, we give only a rough estimate, ignoring the frame difference and the Fermi-blocking in the final state.

The differential luminosities, or energy spectra, of $\nu_e$ and $\bar{\nu}_e$ with the vacuum and MSW neutrino oscillations being taken into account in the adiabatic limit are given as follows:
\begin{eqnarray}
\left( \frac{dL_{N}^{\nu_e}}{dE_{\nu_e}}\right)_{\rm osc} 
&=& p \left( \frac{dL_{N}^{\nu_e}}{dE_{\nu_e}}\right)_0 
+ (1 - p)\left( \frac{dL_{N}^{\nu_x}}{dE_{\nu_x}}\right)_0, \\
\left( \frac{dL_{N}^{\bar{\nu}_e}}{dE_{\bar{\nu}_e}}\right)_{\rm osc} 
&=& p^\prime \left( \frac{dL_{N}^{\bar{\nu}_e}}{dE_{\bar{\nu}_e}}\right)_0 
+ (1 - p^\prime)\left( \frac{dL_{N}^{\bar{\nu}_x}}{dE_{\bar{\nu}_x}}\right)_0. 
\end{eqnarray}
In these expressions, the subscript $0$ means the original spectra before the neutrino oscillations are considered; 
$\nu_{x}$ stands for $\nu_{\mu}$ or $\nu_{\tau}$, both of which we assume are produced solely by electron-positron pair annihilations and have the same spectrum.


\section{Results}

In the following we present the main results: the number luminosities as well as the energy spectra for different neutrino flavors as functions of time. 
Based on them, we then estimate the expected numbers of detection events for different terrestrial neutrino detectors. 

\begin{figure}
\epsscale{1.2}
\plotone{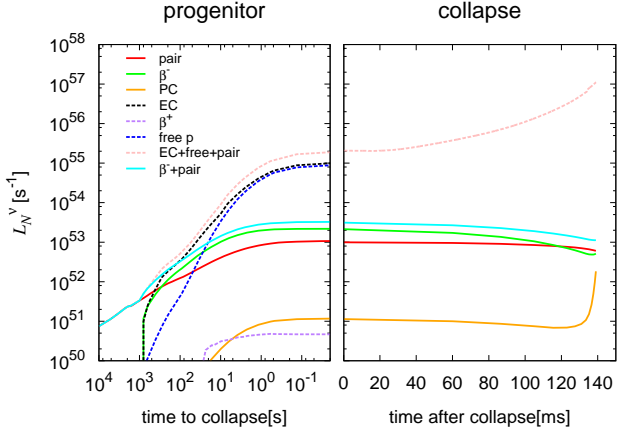}
\caption{The time evolution of neutrino number luminosity for the $15\ M_\odot$ progenitor model. The origin of the horizontal axis corresponds to the time, at which the dynamical simulation is started. Dotted and solid lines show the results for the electron-type neutrino and anti-neutrino, respectively. Colors distinguish the different reactions. In the collapse phase, only total luminosity is shown for $\nu_e$ (pink dotted), since it is the quantity the dynamical simulation provides. Note that the same number of $\nu_e$ and $\bar{\nu}_e$ is produced from the electron-positron pair annihilations (red solid). \label{fig3}}
\end{figure}

\begin{figure}
\epsscale{1.2}
\plotone{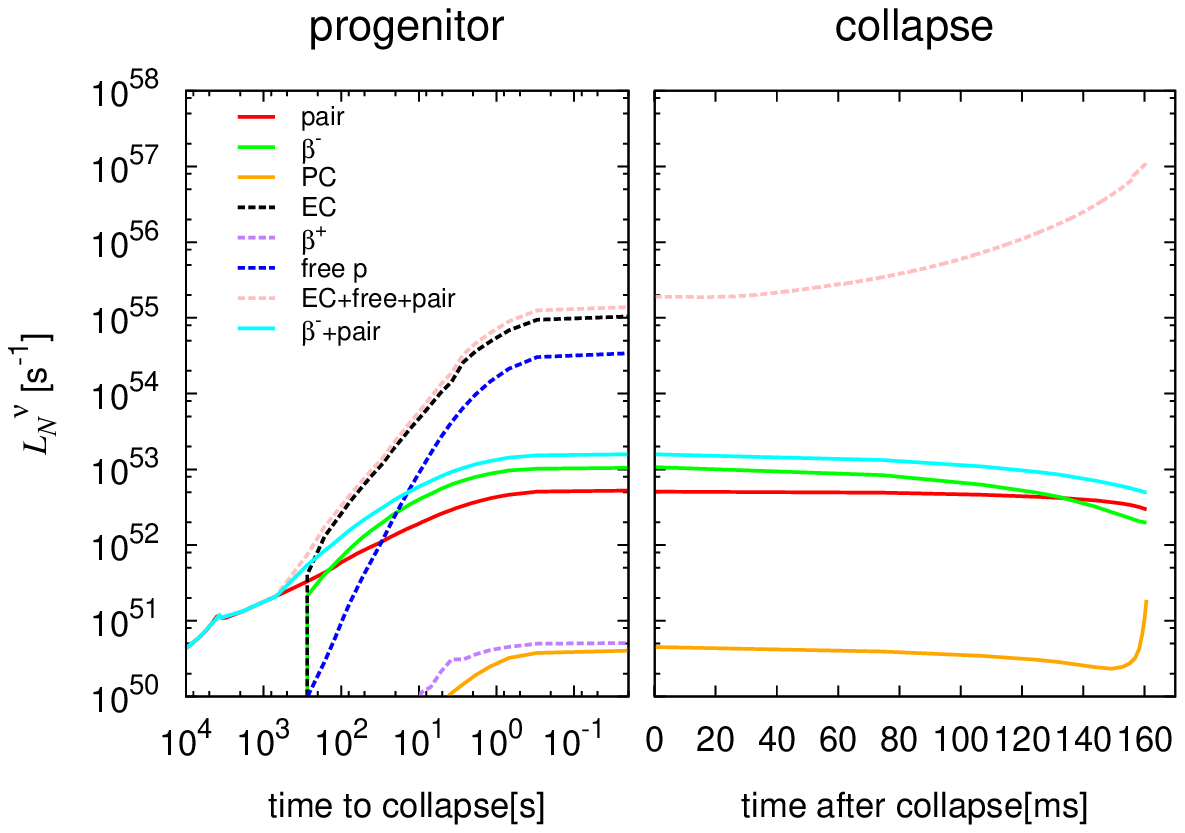}
\caption{The same as Fig.~\ref{fig3} but for the 12 $M_\odot$ progenitor model.\label{fig4}}
\end{figure}

\begin{figure}
\epsscale{1.2}
\plotone{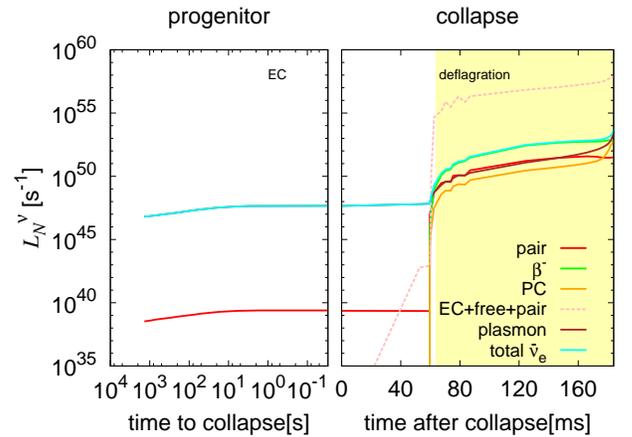}
\caption{The same as Figs.~\ref{fig3} and \ref{fig4} but for the 9 $M_\odot$ progenitor model. \textcolor{black}{The region painted in yellow corresponds to the phase, in which the O+Ne deflagration takes place.} \label{fig5}}
\end{figure}

\subsection{Luminosity and spectrum}
In Fig.~\ref{fig3}, we show the time evolutions of the number luminosities of $\nu_e$ and $\bar{\nu}_e$ for the 15 $M_\odot$ progenitor model.
The left and right panels display the progenitor and collapse phases, respectively.
The origin of the time coordinate corresponds to the time, at which the hydrodynamical calculations are initiated.
The solid and dashed lines denote $\bar{\nu}_e$ and $\nu_e$, respectively.
The colors of lines indicate the contributions from different processes as shown in the legend. 
Note that for $\nu_e$ in the collapse phase we show only the total luminosity, since it is all that the transport calculations produce. 
The nuclear weak processes are considered in the NSE regions alone and, as a result, they arise only after the temperature becomes $T \gtrsim 5 \times 10^9$ K.
It is found that EC's on heavy nuclei and free protons are dominant in the emissions of $\nu_e$ during the progenitor phase,
while the emissions of $\bar{\nu}_e$ occur mainly via the electron-positron pair annihilation until around a few hundreds of seconds before collapse and thereafter the $\beta^-$ decay dominates, which is a new finding in this paper. 
Although $\nu_e$ overwhelms $\bar{\nu}_e$ in the collapse phase as expected, this is also true in the progenitor phase. 
It is particularly the case at $\sim$100 seconds prior to collapse when the electron captures on free protons become appreciable.

Figure \ref{fig4} is the same as Fig.~\ref{fig3} but for the 12 $M_\odot$ progenitor model.
The results are similar to those of the 15 $M_\odot$ model except that the numbers of emitted $\nu_e$ and $\bar{\nu}_e$ are slightly smaller for the 12 $M_\odot$ model than for the 15 $M_\odot$ model because the Fe-core of the 12 $M_\odot$ model has slightly high densities and low temperatures compared to the 15 $M_\odot$ model (see Fig.~\ref{fig2}).

Figure \ref{fig5} shows, on the other hand, the temporal evolutions of the number luminosities in the 9 $M_\odot$ progenitor model, in which the ONe-core collapses to produce an ECSN.
The strong degeneracy of electrons suppresses the electron-positron annihilation in this case and, as a result, the plasmon decay dominates initially until 60 ms after we switch to the hydrodynamical simulation when Ne and O are ignited at the center and the deflagration wave starts to propagate outward to produce NSE behind.
\textcolor{black}{The region painted in yellow corresponds to this O+Ne deflagration phase in the figure.}
Then, $\bar{\nu}_e$ emissions by the $\beta^-$ decay and $\nu_e$ emissions via the EC's on heavy nuclei as well as on free protons overtake those through the plasmon decay.

\begin{figure}
\epsscale{1.0}
\plotone{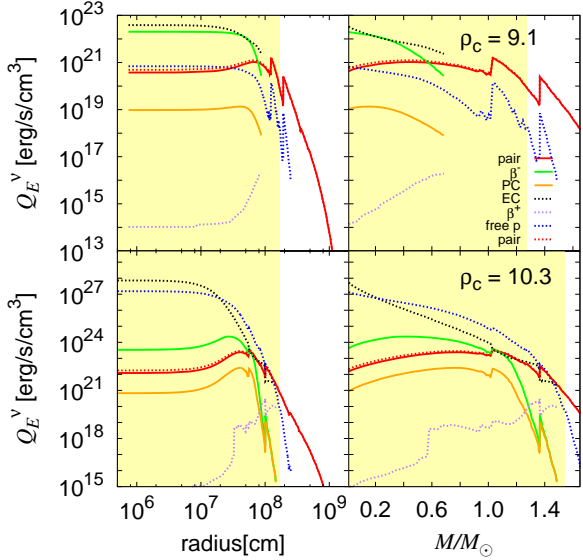}
\caption{The radial profiles of the energy emissivities from different processes for the $15\ M_\odot$ progenitor model. Top and bottom panels show the results when the central density is $\log_{10}{\rho_c}/[\mathrm{g\ cm^{-3}}] = 9.1$ and $10.3$, respectively. In the left panels the radius is used as the horizontal axis whereas in the right panels the mass coordinate is employed.
 The line types and color coding are the same as in Fig.~\ref{fig3}. We define the Fe-cores as the regions, where the electron fraction satisfies $Y_e < 0.495$, and they are painted in yellow in this figure.\label{fig6}}
\end{figure}

In Fig.~\ref{fig6} we present the radial profiles of the energy emissivities, $Q_E^\nu$, from different processes for the 15 $M_\odot$ progenitor model at different times before collapse.
The top panels display the results at a very early time in the progenitor phase ($\log_{10}{\rho_c}/[\mathrm{g\ cm^{-3}}] = 9.1$), with both the radius (left) and mass coordinate (right) being employed as the horizontal axis. We define the Fe-core as the region, where the electron fraction satisfies $Y_e < 0.495$, and paint it in yellow.
It is seen that all emissions occur rather uniformly in the region, $r \lesssim 2\times10^7$ cm, in this early phase.
As the density increases with time, the $\bar{\nu}_e$ emissions are all suppressed toward the center and the peaks in the emissivities appear off center and are shifted to the peripheral, $r \sim5\times10^7$ cm, as shown in the bottom panels of the figure, which correspond to a later time ($\log_{10}{\rho_c}/[\mathrm{g\ cm^{-3}}] = 10.3$ ). This is both due to the depletion of positrons in the initial state and to the Fermi-blocking of electrons in the final state as a consequence of the electron degeneracy. As for $\nu_e$ emissions, such a suppression does not occur and the emissivities are greatest in the central region.

\begin{figure*}
\epsscale{0.8}
\plotone{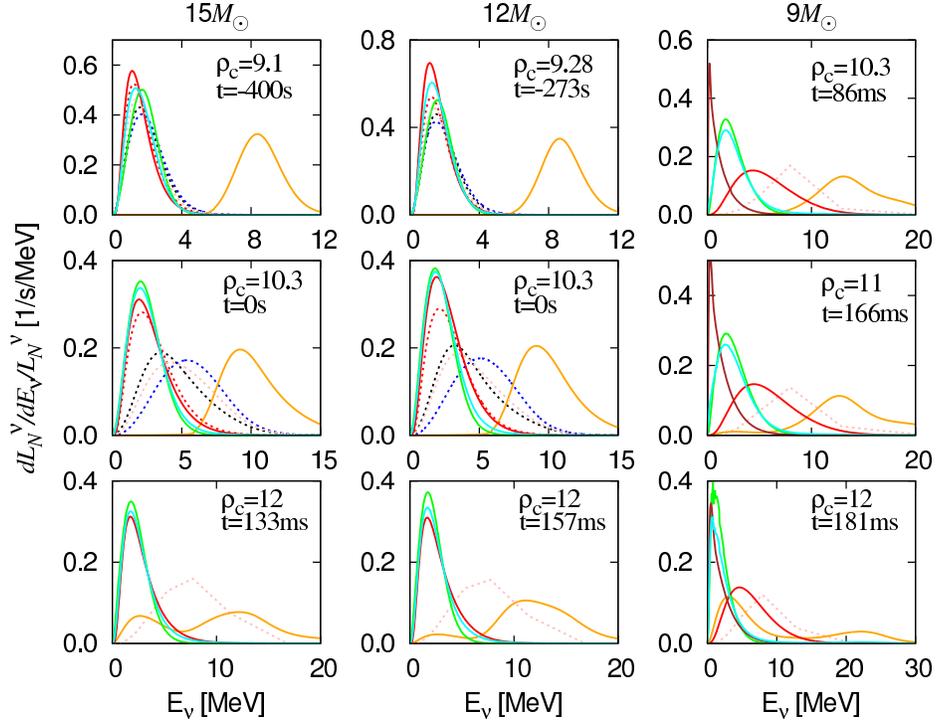}
\caption{The spectra of neutrinos emitted from the entire star at indicated times for the three progenitor models. They are normalized by the corresponding number luminosities. Colors indicate different emission processes as in Fig.~\ref{fig3}. Note that the scale of the horizontal axes are different among the three models.\label{fig7}}
\end{figure*}

Figure \ref{fig7} exhibits the differential luminosities or the energy spectra normalized by the corresponding total luminosities.
The colors and types of lines are the same as those in Fig.~\ref{fig3}.
One can see that $\bar{\nu}_e$'s emitted via PC on heavy nuclei (orange solid lines) have the highest average energies at all times. 
Recall, however, that the luminosity is very low for this process (see Figs.~\ref{fig3}-\ref{fig5}). 
It should be also mentioned that the transport is not solved for $\bar{\nu}_e$, which will not be justified at high densities ($\log_{10}{\rho_c}/[\mathrm{g\ cm^{-3}}] \gtrsim 11$) for these high-energy $\bar{\nu}_e$'s. 
Regardless, the dominant process in the $\bar{\nu}_e$ emission is either the electron-positron annihilation or the $\beta^-$ decay and they both have average energies of 2-5 MeV at most, which may justify the neglect of transport. 
As for the $\nu_e$ emission, the EC's on heavy nuclei and free protons are mostly dominant and produce $\nu_e$'s with $\sim$10 MeV. 
In this case the transport in the core should be computed for the quantitative estimate of the luminosity and spectrum. 
A comparison between the results for the two types of progenitors indicates that neutrinos emitted from the ONe-core progenitor, especially those generated via the electron-positron annihilation, have higher energies than those from the Fe-core progenitors.
This is because electrons are more strongly degenerate and have greater chemical potentials in the former.


In Table \ref{tab3}, we list the top five contributors to the EC and $\beta^-$ decay, the dominant processes to produce $\nu_e$ and $\bar{\nu}_e$, respectively, at the time when $\log_{10}{\rho_c}/[\mathrm{g\ cm^{-3}}] = 10.3$ in the 15 $M_\odot$ model. 
Note that the EC occurs mostly in the central region whereas the $\beta^-$ decay happens off center mainly. 
We hence evaluate the EC rates at $r=3.1\times 10^5\ \mathrm{cm}$, where the density, temperature and electron fraction are $\log_{10}{\rho}/[\mathrm{g\ cm^{-3}}] = 10.3$, $T=0.861$ MeV, $Y_e = 0.417$ and $\mu_e = 11.9$ MeV. The $\beta^-$ decay rates are presented, on the other hand, for the condition at $r=2.7\times 10^7$ cm, i.e., $\log_{10}{\rho}/[\mathrm{g\ cm^{-3}}]=9.79$, $T=0.856$ MeV, $Y_e=0.423$ and $\mu_e = 7.87$ MeV. 
We find that although the emissivities for individual nuclei are proportional to the product of their mass fraction and the reaction rate, the former is more important, since the latter changes by a factor whereas the former varies by an order. 
It is noteworthy in this respect that the top two contributors to the EC and the top one to the $\beta^-$ decay are those nuclei with magic proton numbers, which is the reason why they are more abundant than others. Note again that their reaction rates are not the greatest.

\begin{figure}
\epsscale{1.2}
\plotone{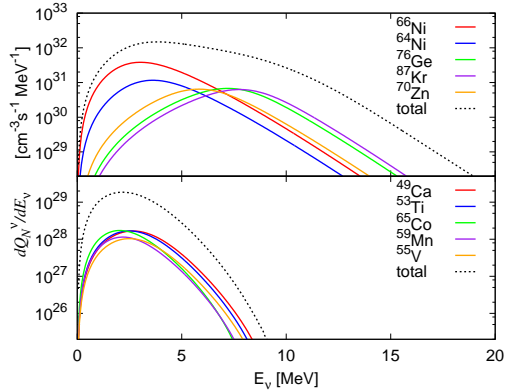}
\caption{The energy spectra for the EC and $\beta^-$ decay by the dominant heavy nuclei given in Table \ref{tab3} at the time when the central density becomes $\log_{10}{\rho_c}/[\mathrm{g\ cm^{-3}}] = 10.3$.
The top panel shows the $\nu_e$ spectrum for EC, while the bottom panel displays the $\bar{\nu}_e$ spectrum for $\beta^-$ decay. The EC rates are evaluated at $r=3.1 \times 10^5$ cm, where the density, temperature and electron fraction are $\log_{10}{\rho}/[\mathrm{g\ cm^{-3}}] = 10.3$, $T=0.861$ MeV, $Y_e = 0.417$, $\mu_e=11.9$ MeV respectively. The rates of $\beta^-$ decays are calculated at $r=2.7 \times 10^7$ cm, where they are largest and $\log_{10}{\rho}/[\mathrm{g\ cm^{-3}}] = 9.79$, $T=0.856$ MeV, $Y_e = 0.423$, $\mu_e=7.87$ MeV. \label{fig8}}
\end{figure}

In Fig.~\ref{fig8}, we show the energy spectra of neutrinos emitted from these nuclei. 
It is recognized that the spectra for the $\beta^-$ decay presented in the lower panel are not much different among the nuclei. 
It is also evident that the average energies are lower than those for the $\nu_e$'s emitted through the EC's as exhibited in the upper panel.
This is because the latter includes the contribution from the kinetic energy of degenerate electrons. 
The variation among the nuclei is also larger for the EC. 

\textcolor{black}{Once the NSE is established after the passage of the deflagration wave in the 9 $M_\odot$ model, the composition is simply determined by the density, temperature and electron fraction. 
The iron-group elements hence become dominant for EC and $\beta^-$ decay also in the 9$M_\odot$ model just as in the 12 and 15$M_\odot$ models.}

\begin{table*}
\begin{center}
\caption{Weak reaction rates and mass fractions of the top five nuclei contributing to the total number luminosities from EC and $\beta$ decay in the 15 $M_\odot$ progenitor model at the time when the central density is $\log_{10}{\rho_c}/[\mathrm{g\ cm^{-3}}] = 10.3$.\tablenotemark{a} \label{tab3}}
\begin{tabular}{cccc|ccccccccc}
\tableline\tableline
\multicolumn{4}{c|}{EC\tablenotemark{b}}  & \multicolumn{4}{c}{$\beta^-$\tablenotemark{c}} \\ \hline
 & $(N,Z)$ & $X_i$ & $R_i$ &  & $(N,Z)$ & $X_i$ & $R_i$ \\ \hline
\tableline
${}^{66}$Ni &(38,28)  & $7.76\times10^{-2}$ & 10.57 &${}^{49}$Ca& (29,20) & $1.88\times10^{-2}$ & $3.64\times10^{-2}$ \\
${}^{64}$Ni &(36,28)  & $1.99\times10^{-2}$ & 11.89 &${}^{53}$Ti& (31,22) & $1.29\times10^{-2}$ & $5.56\times10^{-2}$ \\
${}^{76}$Ge &(44,32)  & $5.88\times10^{-3}$ & 32.59 &${}^{65}$Co& (38,27) & $4.60\times10^{-3}$ & $1.78\times10^{-1}$ \\
${}^{87}$Kr &(51,36)  & $7.85\times10^{-3}$ & 26.37 &${}^{59}$Mn& (34,25) & $9.78\times10^{-3}$ & $5.20\times10^{-2}$ \\
${}^{70}$Zn &(40,30)  & $5.32\times10^{-3}$ & 30.04 &${}^{55}$V & (32,23) & $6.05\times10^{-3}$ & $7.62\times10^{-2}$ \\
\tableline
\end{tabular}
\end{center}
\begin{flushleft}
\tablenotetext{a}{This density corresponds to the time, at which we switch to the dynamical calculation ($t = 0$).}
\tablenotetext{b}{The EC rates are evaluated at $r=3.1 \times 10^5$ cm, where the density, temperature and electron fraction are $\log_{10}{\rho}/[\mathrm{g\ cm^{-3}}] = 10.3$, $T=0.861$ MeV, $Y_e = 0.417$, $\mu_e=11.9$ MeV. }
\tablenotetext{c}{The rates of $\beta^-$ decays are calculated at $r=2.7 \times 10^7$ cm, where they are largest and $\log_{10}{\rho}/[\mathrm{g\ cm^{-3}}] = 9.79$, $T=0.856$ MeV, $Y_e = 0.423$, $\mu_e=7.87$ MeV.}
\end{flushleft}
\end{table*}

\subsection{Event numbers at detectors}
Based on the results obtained so far, we estimate the numbers of detection events for some representative detectors, which include those under planning at present.
For the detection of $\bar{\nu}_e$, almost all detectors utilize the inverse $\beta$ decay:
\begin{equation}
\bar{\nu}_e + p \longrightarrow e^+ + n.
\end{equation}
Following \cite{odrzy04}, we express the cross section $\sigma(E_\nu)$ of this interaction as 
\begin{equation}
\sigma(E_\nu)=0.0952\left(\frac{E_{e^+}p_{e^+}}{1\mathrm{MeV}^2}\right)\times 10^{-42} \ \ \ \ \mathrm{cm}^2,
\end{equation}
in which the energy and 3-momentum of the positron emitted are denoted by $E_{e^+}=E_{\nu}-\left(m_n-m_p\right)$ and $p_{e^+}=\sqrt{{E_{e^+}}^2-{m_e}^2}$, respectively.

Electron neutrinos are normally detected via the electron-scattering: $\nu_e + e^- \longrightarrow \nu_e + e^-$, in the currently available detectors.
Its reaction rate is much lower than that of the inverse $\beta$ decay, however, and the detection of $\nu_e$'s in the pre-bounce phase has been thought to be almost impossible.
Then the new-type detector using liquid Argon has come into view.
The planned deep underground neutrino observatory, or DUNE, is one of the such detectors \citep{dune}.
It employs the absorption of $\nu_e$ by $^{40}\rm{Ar}$:
\begin{equation}
^{40}\mathrm{Ar} + \nu_e \longrightarrow e^- + ^{40}\mathrm{K^\ast}.
\end{equation}
The cross section of this reaction is obtained numerically with \citeauthor{scholberg}.

Then the event rate, $r$, at a detector is expressed as
\begin{equation}
r=\frac{N}{4\pi D^2}\int_{E_{\mathrm{th}}}^{\infty} dE_{\nu_1} \sigma\left(E_{\nu_1}\right) \frac{dL_{N}^{\nu_1}}{dE_{\nu_1}}, 
\label{eq:eq42}
\end{equation}
in which $N$ and \textcolor{black}{$D$} denote the target number in the detector and the distance to the star from the detector, respectively. 
For simplicity, we assume that the detection efficiency is $100\ \%$ above the threshold energy $E_{\rm th}$. 
The features of relevance for the detectors that we consider in this paper, i.e, Super-Kamiokande, KamLAND, Hyper-Kamiokande, JUNO and DUNE, are summarized in Table~\ref{detector}. 
The cumulative number of events, $N_{\rm cum}$, is obtained by integrating the rate up to the given time:
\begin{equation}
N_{\rm cum}(t) = \int_{t_{\rm ini}}^t r \, dt.
\end{equation}

\begin{figure}
\epsscale{2.0}
\plottwo{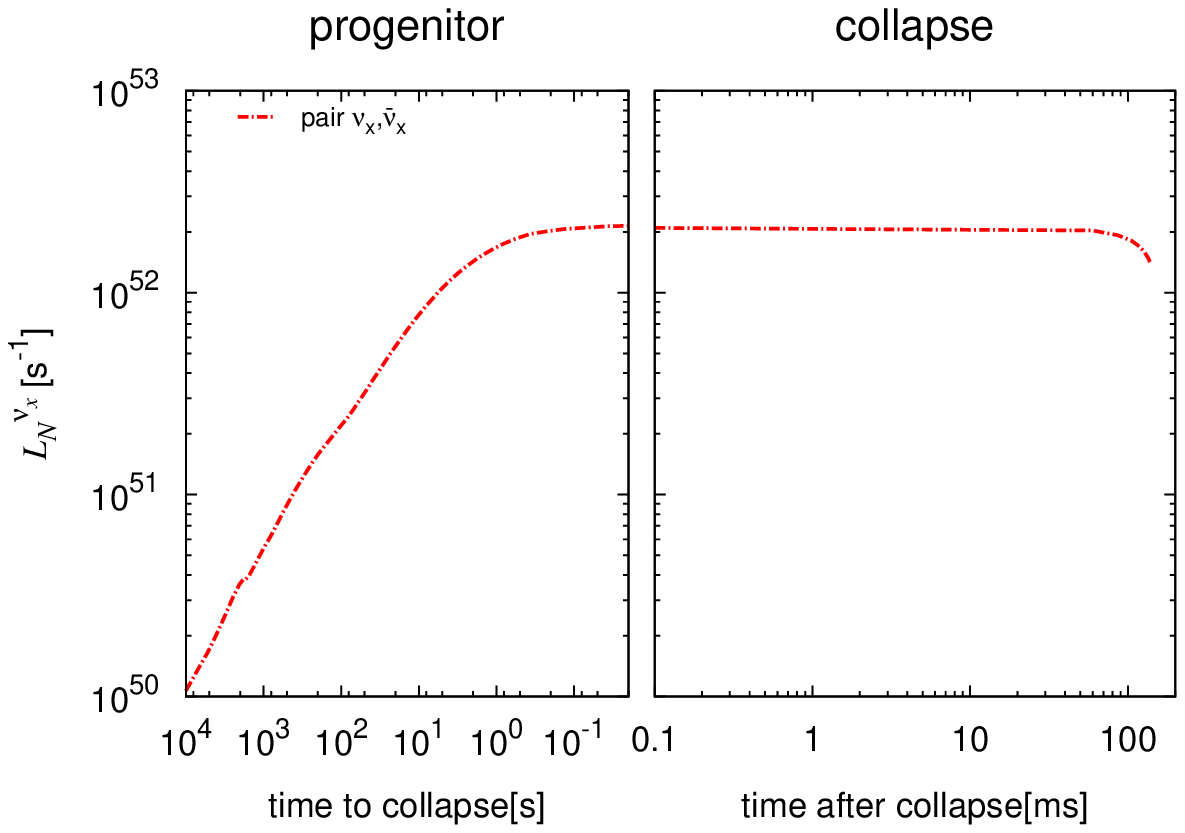}{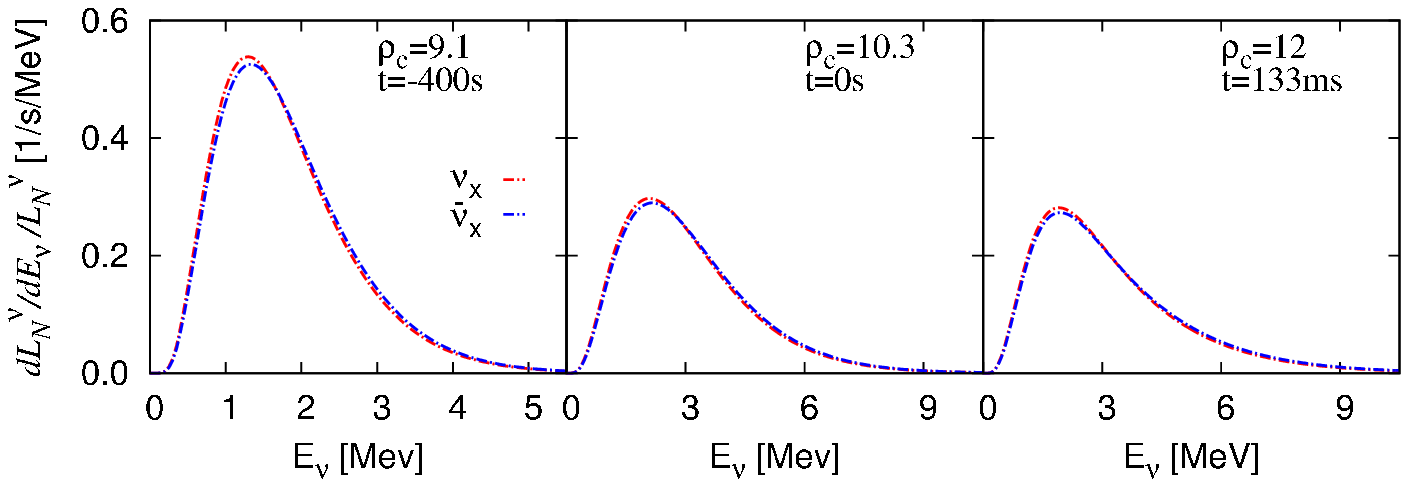}
\caption{The number luminosity (top) and normalized spectra (bottom) of $\nu_x$ and $\bar{\nu}_x$ emitted via the electron-positron annihilation. Note that the number luminosities of $\nu_x$ and $\bar{\nu}_x$ are identical to each other. \label{fignux}}
\end{figure}

In order to give quantitative estimates to  the numbers of detection events, we need to take into account neutrino oscillations appropriately. 
For that purpose, not only the luminosities of electron-type neutrinos but also those of mu- and tau-types neutrinos are required. 
In this paper we have calculated them for the electron-positron annihilation on the same basis as $\nu_e$ and $\bar{\nu}_e$. 
We give the results in Fig.~\ref{fignux}, in which the time evolution of the number luminosities as well as the energy spectra at three different epochs are displayed in the upper and lower panels, respectively. 
It is observed that the luminosities are much smaller than those of $\nu_e$ as expected and are somewhat lower even compared with $\bar{\nu}_e$. 
This is simply because that $\mu$- and $\tau$-types neutrinos lack charged-current reactions and are produced solely from the electron-positron annihilation. 
The average energies are $\lesssim$ 2 MeV, much lower than that of $\nu_e$ and, as a result, the opacities for these heavy-lepton neutrinos are smaller, justifying the neglect of transport in their calculations.

\begin{table*}
\begin{center}
\caption{The detector parameters assumed in this paper.\tablenotemark{a,b,c}\label{detector}}
\begin{tabular}{cccccc}
\tableline\tableline
Detector &$\ \ \ \ \ \ \ $ Mass &$\ \ \ \ \ \ \ $ Target number &$\ \ \ \ \ \ \ $ Energy threshold  \\
&$\ \ \ \ \ \ \ $[kt]&$\ \ \ \ \ \ \ $N&$\ \ \ \ \ \ \ $[MeV] \\
\tableline
Super-K & $\ \ \ \ \ \ \ $32 &$\ \ \ \ \ \ \ $2.14$\times 10^{33}$ &$\ \ \ \ \ \ \ $5.3 \\
KamLAND&$\ \ \ \ \ \ \ $1 &$\ \ \ \ \ \ \ $8.47$\times 10^{31}$ &$\ \ \ \ \ \ $ 1.8\\
Hyper-K &$\ \ \ \ \ \ \ $516  &$\ \ \ \ \ \ \ $3.45$\times 10^{34}$ &$\ \ \ \ \ \ \ $8.3 & \\
JUNO&$\ \ \ \ \ \ \ $20&$\ \ \ \ \ \ \ $1.69$\times 10^{33}$&$\ \ \ \ \ \ \ $1.8\\
DUNE&$\ \ \ \ \ \ \ $40&$\ \ \ \ \ \ \ $6.02$\times 10^{32}$&$\ \ \ \ \ \ \ $5.0, 10.8\\
\tableline
\end{tabular}
\end{center}
\begin{flushleft}
\tablenotetext{a}{The numbers given here are not very precise and just meant for a rough estimate. JUNO is assumed to be a scale-up of KamLAND by a factor of 20. We also assume that the energy threshold of Hyper-Kamiokande will be somewhat higher than that of Super-Kamiokande.}
\tablenotetext{b}{We use the total volume for the 2 tank-design of Hyper-Kamiokande.}
\tablenotetext{c}{The energy threshold of DUNE is still uncertain and we employ both an optimistic (5 MeV) and more realistic (10.8 MeV) values in this study. }
\tablerefs{
(1) \citealt{sk}; (2) \citealt{kam}; (3) \citealt{hk}; (4) \citealt{juno14}; (5) \citealt{dune}}
\end{flushleft}
\end{table*}

\begin{figure}
\epsscale{2.5}
\plottwo{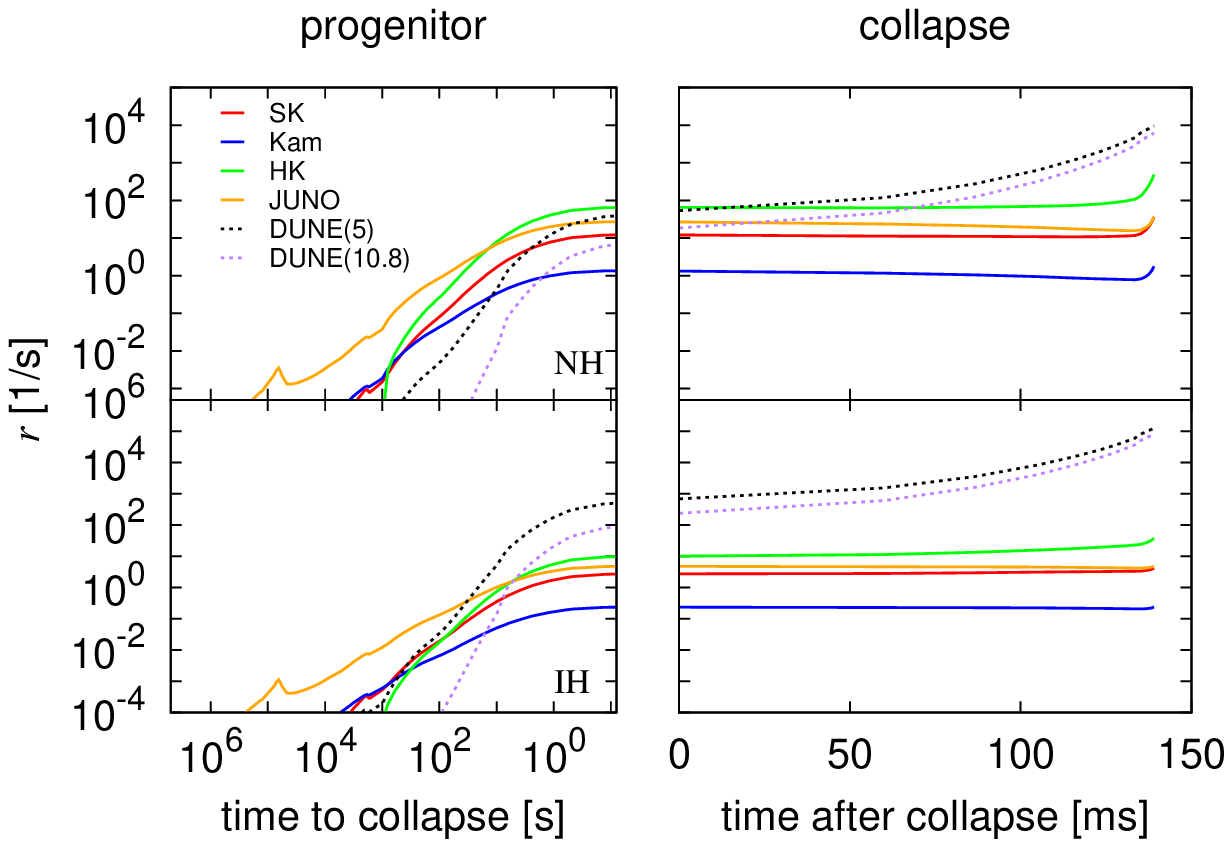}{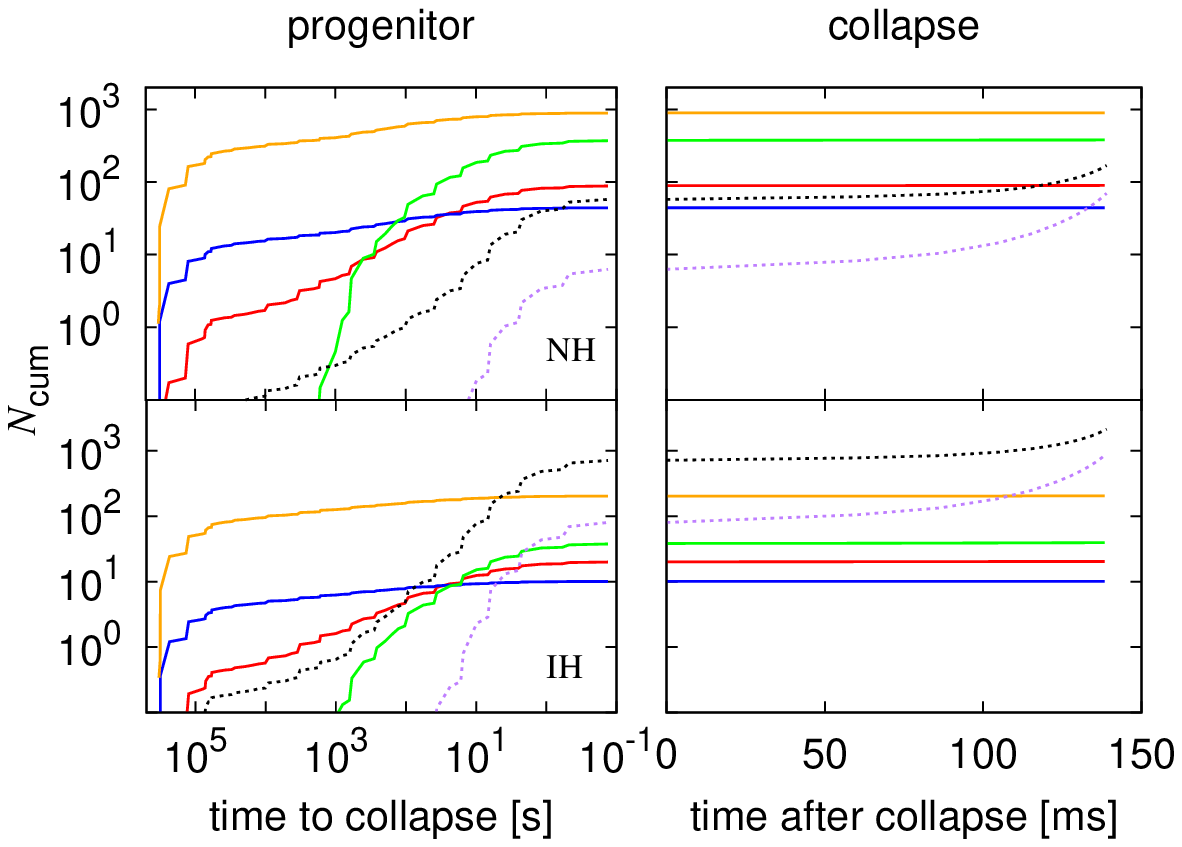}
\caption{The time evolutions of event rates (top panels) and the cumulative numbers of events (bottom panels) for the 15 $M_\odot$ progenitor model. The upper half of each panel shows the results for the normal mass hierarchy, while the lower half presents those for the inverted mass hierarchy. Colors specify neutrino detectors. We consider $\nu_e$ for DUNE (dotted line) and $\bar{\nu}_e$ for other detectors (solid lines).  \label{fig10}}
\end{figure}

\begin{figure}
\epsscale{2.5}
\plottwo{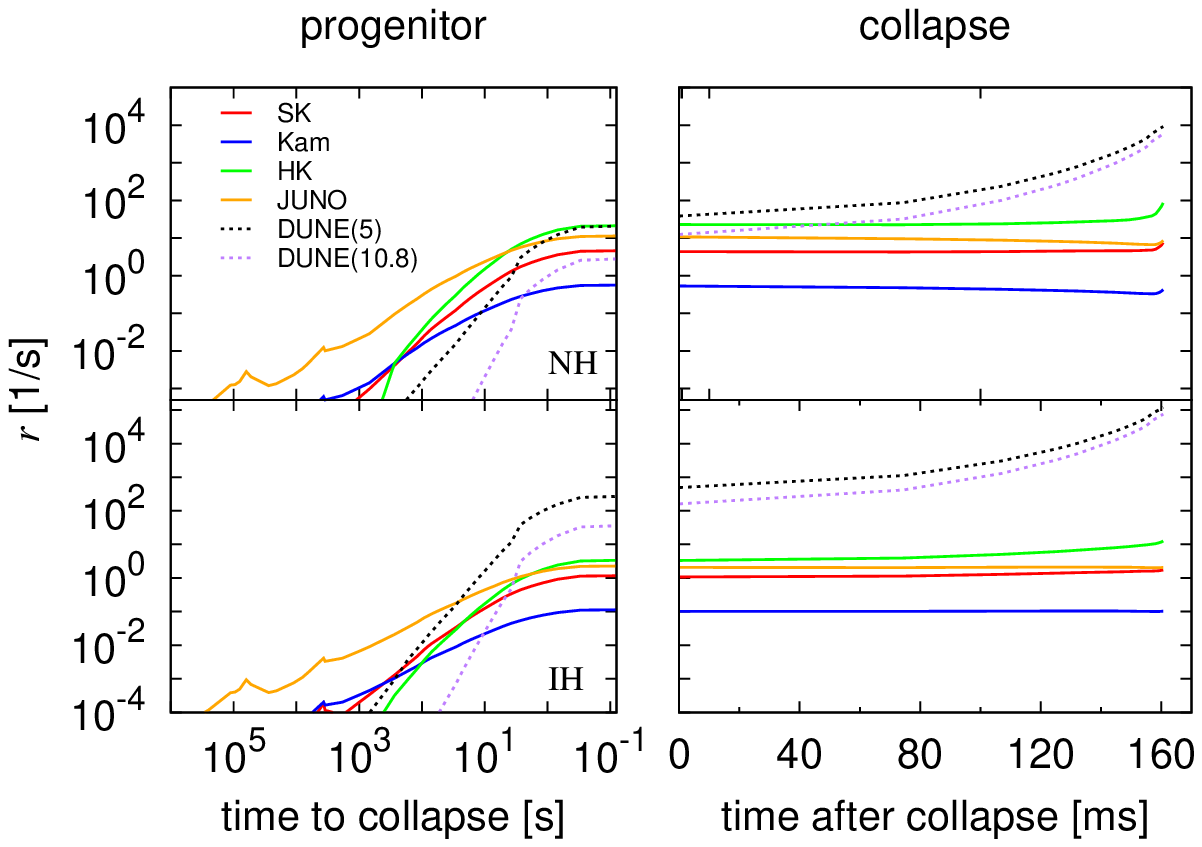}{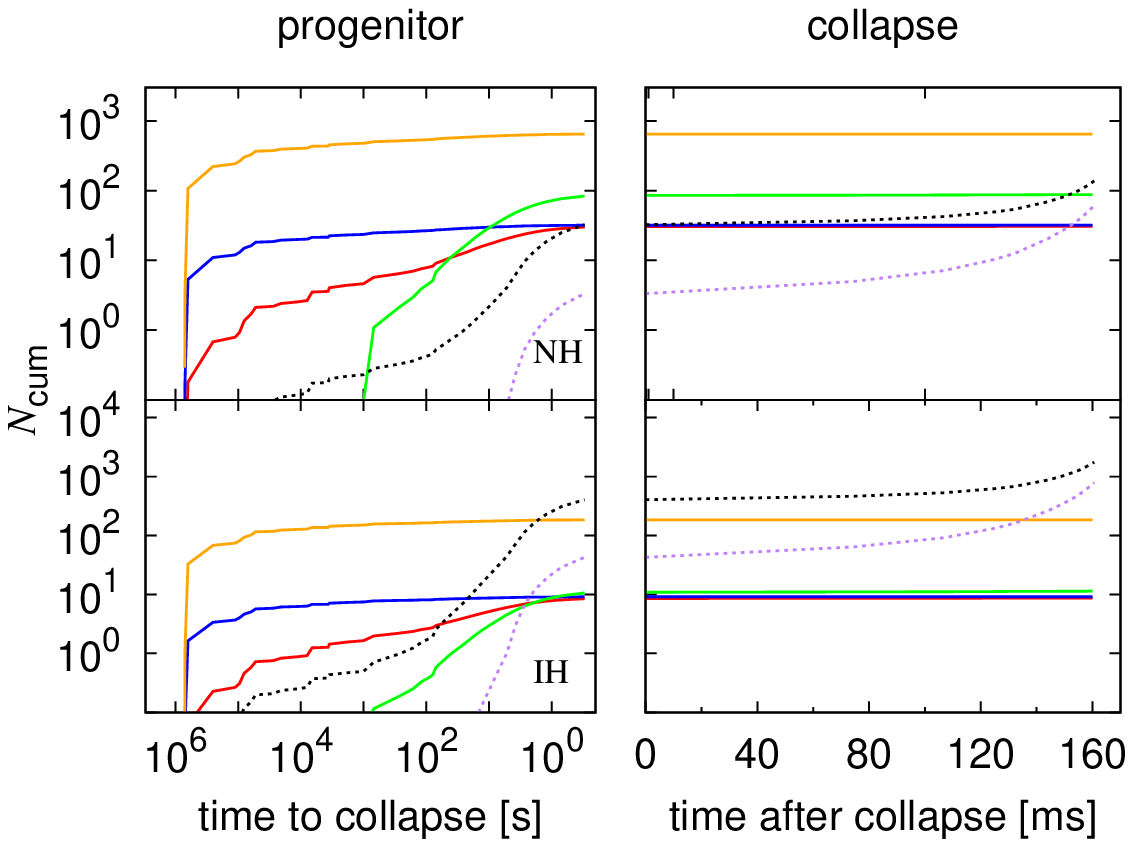}
\caption{The same as Fig.~\ref{fig10} but for the 12 $M_\odot$ progenitor. \label{fig11}}
\end{figure}

\begin{figure}
\epsscale{2.5}
\plottwo{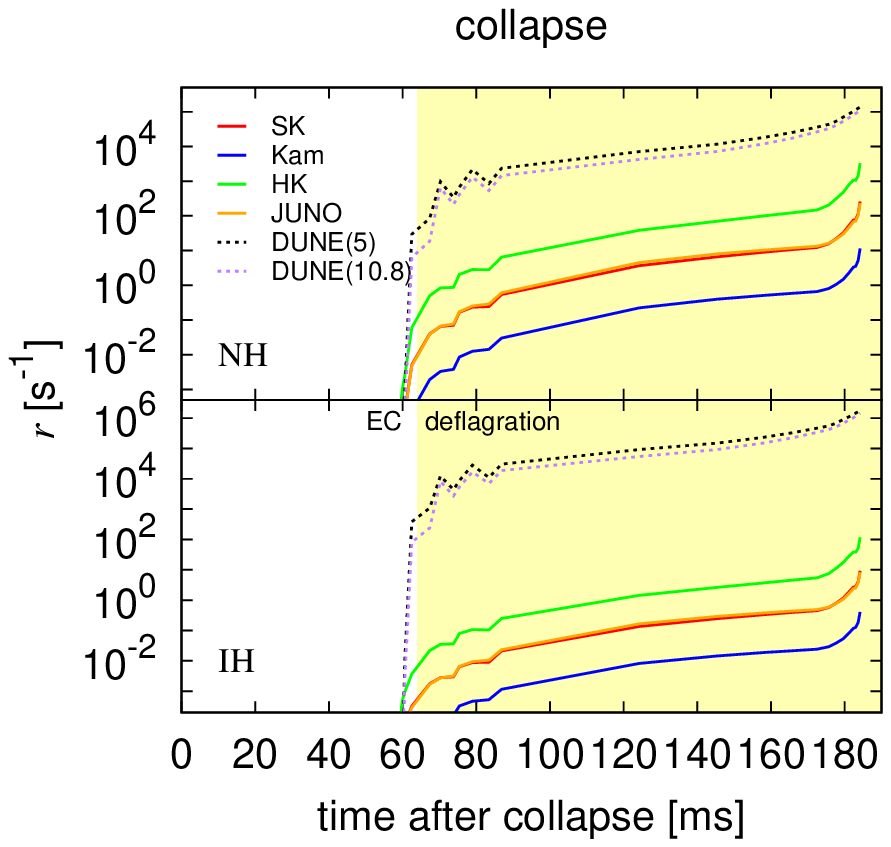}{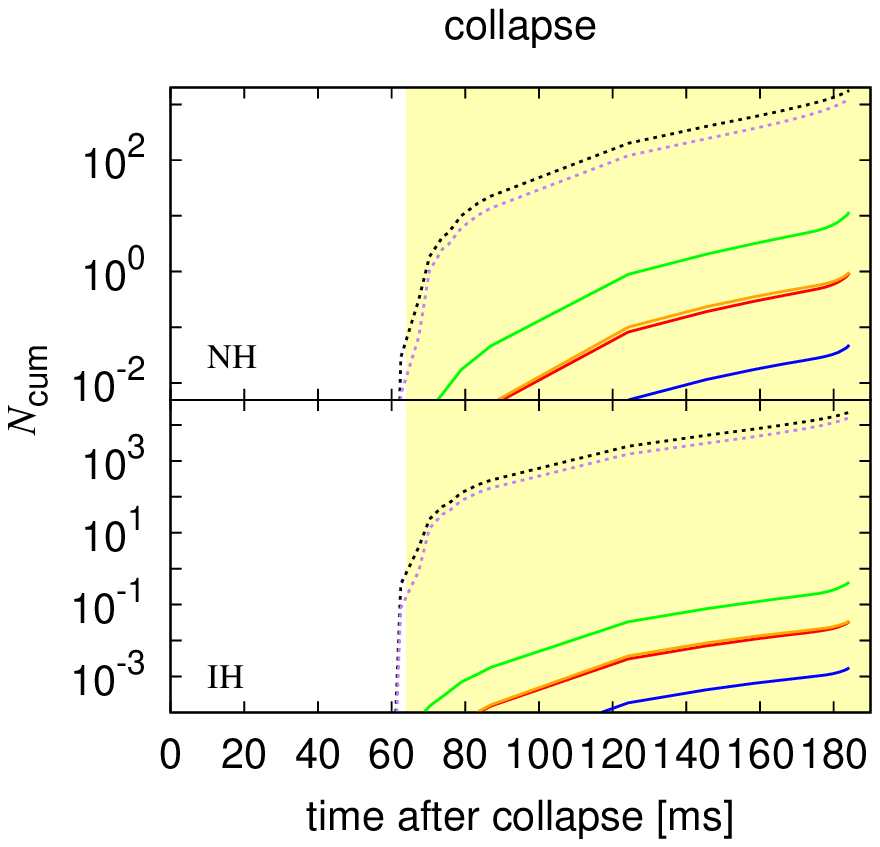}
\caption{The same as Figs.~\ref{fig10} and \ref{fig11} but for the 9 $M_\odot$ progenitor. Only the collapse phase is shown. \label{fig12}}
\end{figure}

Figures \ref{fig10}, \ref{fig11} and \ref{fig12} present the time evolutions of event rates (top) and cumulative numbers of detection events (bottom) for different detectors in the progenitor (left) and collapse (right) phases for the three progenitors. 
For the 9 $M_\odot$ model, only the collapse phase is shown, since the progenitor phase will not be observed even at the distance as close as 200 pc (\textcolor{black}{Paper I}). 
The normal (inverted) hierarchy is assumed in the upper (lower) half of the top panels in each figure. 
All the detectors except DUNE, which will detect $\nu_e$'s, will observe $\bar{\nu}_e$'s mainly. 
From the comparison of the left and right panels we find that the progenitor phase is dominant over the collapse phase for $\bar{\nu}_e$ with the latter contributing only a few percent.
This is due to the electron degeneracy, which suppresses both the $\beta^-$ decay via the Fermi-blocking of the electron in the final state and the electron-positron annihilation through the depletion of the positron in the initial state. 

In the case of $\nu_e$, the collapse phase is much more important although it lasts for much shorter periods. 
This is because both the luminosity and the average energy rise with the density.
The detections of $\bar{\nu}_e$'s in the pre-bounce phase are hence more suitable for the alert of the imminent supernova \citep{asakura16,yoshida16}. 
In fact, we may be able to issue an alert a few days before the core collapse for the Fe-core progenitors if neutrinos obey the normal mass hierarchy.
The $\nu_e$ emissions from the ONe-core progenitor, on the other hand, are much shorter than those from the Fe-core progenitors presented in Fig.~\ref{fig12}.
They become appreciable only after NSE is established in the collapsing core by the passage of the deflagration wave. 
DUNE will detect $\nu_e$'s only from less than 100 ms prior to bounce, and may be hence possible to distinguish the two types of progenitors by the time of the first detection of $\nu_e$'s.

Depending on the mass hierarchy, the neutrino oscillations affect either $\nu_e$ or $\bar{\nu}_e$ predominantly. 
In fact, in the normal hierarchy, the spectrum of $\nu_e$ is exchanged with that of $\nu_\tau$ in the adiabatic MSW oscillation and is further mixed among three flavors in the vacuum oscillations whereas the spectrum of $\bar{\nu}_e$ is mixed with those of $\bar{\nu}_\mu$ and $\bar{\nu}_\tau$ only in the vacuum oscillations. 
The situation is the other way around in the case of the inverted hierarchy, in which the MSW affects $\bar{\nu}_e$ also. 
It is recalled that the luminosities of $\nu_x$ and $\bar{\nu}_x$ are lower than those of $\nu_e$ and $\bar{\nu}_e$. 

As a consequence, the chance to observe $\bar{\nu}_e$'s is higher for the normal hierarchy and JUNO will see more than 850 of them in the progenitor phase from as early as a few days prior to collapse, which is roughly the end of O burning, if the 15 $M_\odot$ progenitor is located at 200 pc from the earth. 
The event number will be reduced by a factor of $\sim$4 in the case of the inverted hierarchy. 
The detection of $\nu_e$'s on DUNE will be more plausible for the inverted hierarchy and, in fact, the expected event number may exceed 2000 if the distance to the source is again 200 pc, i.e., the distance to Betelgeuse and the energy threshold is optimistically assumed to be 5 MeV. 
The first $\nu_e$ may be observed several tens of minutes before collapse, which corresponds to the end of Si-burning.
In the normal hierarchy, on the hand, the number of detections will be reduced by more than a factor of 10.
Such a large difference in the number of detections suggests a possibility to judge the neutrino mass hierarchy.
It is interesting to point out that as far as the $\nu_e$ is concerned, the ONe-core progenitor may offer a better chance of detection at DUNE. 
This is because the temperature in the NSE region behind the O+Ne deflagration is higher than in the Fe-core.
As long as the Fe-core progenitors are concerned, the more massive it is, the larger number of detection events are expected both for $\nu_e$ and $\bar{\nu}_e$.

\begin{table*}
\begin{center}
\caption{The expected numbers of detection events for different detectors.  \tablenotemark{a,b} \label{eventrate}}
\begin{tabular}{ccccccccccc}
\tableline\tableline
detector & \multicolumn{2}{c}{9 $M_\odot$} & \multicolumn{2}{c}{12 $M_\odot$} & \multicolumn{2}{c}{15 $M_\odot$}  \\
&normal&inverted&normal&inverted&normal&inverted \\
\tableline
Super-K & 0.93 & 0.03  &     30.8   &    8.68    & 89.9 & 20.3\\
        &      &       & (30.1+0.71)& (8.48+0.20) & (88.3+1.61) & (19.9+0.41)\\
KamLAND & 0.05 & 0.002 &     32.0   & 9.15 & 44.3 & 10.1 \\
        &      &       & (31.9+0.07) & (9.13+0.02) & (44.2+0.15)& (10.1+0.03)\\ 
Hyper-K & 11.6 & 0.42  & 83.9 & 10.9 & 363 & 37.7\\
        &      &       & (80.0+3.85)& (10.1+0.76) & (353+9.84) & (35.9+1.82)\\ 
JUNO    & 0.98 & 0.04  & 645   & 184  & 894 & 204\\
        &      &       & (644+1.47) & (184+0.33) & (891+3.07) & (203+0.63)\\
DUNE(5MeV)    & 1765  & 22685  & 137 & 1756 & 169 & 2142 \\
        &      &       & (32.4+105) & (406+1350)& (57.8+111)&(713+1429)\\ 
DUNE(10.8MeV) & 1238 & 15910 & 61.3 & 789 & 69.3 & 895 \\
              &    &   & (3.33+58.0) & (42.7+746) & (6.27+63) & (80.1+815) \\
\tableline
\end{tabular}
\tablenotetext{a}{The numbers are pertinent to $\nu_e$ for DUNE and to $\bar{\nu}_e$ for other detectors. In the case of the Fe-core progenitors, the individual contributions from the progenitor and collapse phases are also shown in the parentheses in this order.}
\tablenotetext{b}{The source is assumed to be located at $200\ {\rm pc}$ from the earth. Both the normal and inverted mass hierarchies are considered in the adiabatic oscillation limit.}
\end{center}
\end{table*}

In Table \ref{eventrate}, we summarize the expected numbers of events at Super-Kamiokande, KamLAND, Hyper-Kamiokande, JUNO and DUNE, assuming that progenitors are 200 pc away from us.
They are the numbers of $\nu_e$ events for DUNE and those of $\bar{\nu}_e$ events for other detectors. 
In the table, the contributions from both the progenitor and collapse phases are exhibited.
It is found that $\bar{\nu}_e$'s from the 12 and 15 $M_\odot$ progenitors can be detected at all detectors if the source is this close. In particular, the planned detectors such as Hyper-Kamiokande and JUNO look promising if one considers the number of events alone: they will detect a few tens of $\bar{\nu}_e$'s even if they are emitted from 1kpc away.
The detection of $\bar{\nu}_e$'s from the ONe-core progenitor seems to be nearly impossible even with the planned detectors.
We will be hence able to distinguish the two types of progenitors, i.e., ONe-core progenitors and Fe-core progenitors, by detection or non-detection of $\bar{\nu}_e$, the same conclusion as in \textcolor{black}{Paper I}.
It is stressed, however, that in this paper we have incorporated nuclear processes, such as $\beta^-$ decay, which were neglected in \textcolor{black}{Paper I} but are demonstrated to be dominant in the production of $\bar{\nu}_e$.
We show the expected numbers of $\nu_e$ events for the values of two energy thresholds, 5 and 10.8 MeV, considering its uncertainties at present.
The former is somewhat optimistic and the latter may be more realistic.
The first detection of $\nu_e$ will be delayed for the latter case to a few tens of seconds before core bounce.
Note, however, that the energy of $\nu_e$'s in the collapse phase is high ($\sim$8 MeV) and we will be still able to detect a large number of $\nu_e$'s.
In this paper, we do not treat the neutrino emissions at $\log_{10}{\rho}/[\mathrm{g\ cm^{-3}}] > 13$ because the compositions and weak reaction rates of heavy nuclei are highly uncertain there.
The number of events for $\bar{\nu}_e$ will not increase much by the time of core bounce, however. In fact, it is expected to increase by $\sim$200 for $\nu_e$ if one simply extrapolates the results obtained for $\log_{10}{\rho_c}/[\mathrm{g\ cm^{-3}}] < 13$ up to core bounce in the inverted hierarchy.
This issue will be addressed in a future publication.


\section{summary and discussions}

In our previous paper \citep[Paper I]{kato15}, we calculated $\bar{\nu}_e$ emissions via thermal processes alone: the electron-positron pair annihilation and plasmon decay from both the Fe-core progenitors and the ONe-core progenitor. 
The nuclear weak processes, i.e., the $\beta^{\mp}$ decays and electron- and positron captures, were ignored, however. 
Moreover, the neutrino emissions in the collapse phase were not considered, either, because the computations of hydrodynamics and neutrino transfer would have been required. 
These neglects may no longer be justified as the liquid Ar detector such as DUNE has come into view to detect $\nu_e$'s, which are predominant in the collapse phase but are difficult to observe for the existing detectors. 
It should be stressed here that no quantitative estimate has been done so far on the neutrino emissions during the collapse phase mainly because neutrinos are emitted more intensively in the post-bounce phase and the proto-neutron star cooling phase that follows. 
This paper is hence the first to demonstrate that the collapsing phase has a potential to provide new insights.

In this paper, we have investigated the emissions of all-types of neutrinos from the progenitor phase up to the pre-bounce time, at which the central density reaches $\log_{10}{\rho_c}/[\mathrm{g\ cm^{-3}}] = 13$. 
We have compared the two types of progenitors of CCSNe: one that produces the Fe-core and the other that yields the ONe-core before core collapse, to see whether we can get some information on the cores deep inside massive stars, which would be inaccessible to other means, by observing the neutrinos they emit.
We have first re-calculated the neutrino emissions from the realistic progenitor models with 9, 12 and 15 $M_\odot$ on the zero age main sequence with both the thermal and nuclear weak processes being taken into account. 
\textcolor{black}{Note again that the 9 $M_\odot$ model is a progenitor with the ONe-core that collapses to produce the ECSN and the other two are supposed to be progenitors of the FeCCSNe. }

We have then switched to hydrodynamical simulations of core collapse up to the pre-bounce time, at which the central denisty reaches $\log_{10}\rho_c/\mathrm{[g/cm^3]}=13$, with the transfer of $\nu_e$ in the core being treated appropriately. 
Since other types of neutrinos are much less abundant and have lower energies typically, we have treated them in the post-process, in which we have extracted the time evolutions of density, temperature, $Y_e$ and $f_{\nu_e}$, the distribution function of $\nu_e$, from results of the simulations and calculated the emissivities of these neutrinos with possible minor back-reactions to dynamics being ignored. 
Finally, based on the luminosities and spectra of neutrinos thus obtained, we have estimated the expected numbers of detection events on some representative neutrino detectors. 
In so doing we have taken into proper account the vacuum and MSW oscillations of neutrino flavors.

We have found that the $\beta^-$ decay and the EC on heavy nuclei and free protons dominate the number luminosities of $\bar{\nu}_e$ and $\nu_e$, respectively, from several tens of minutes before core bounce.
To these reactions heavy nuclei not with large reaction rates but with large mass fractions contribute most.
Because of the Fermi-blocking of electrons in the final state, the $\beta^-$ decay is suppressed at high densities, where electrons are strongly degenerate, and the number luminosity of $\bar{\nu}_e$ is decreased toward core bounce.
As a consequence, the progenitor phase is dominant over the collapse phase in the $\bar{\nu}_e$ emission.
In contrast, the $\nu_e$ emission occurs predominantly in the collapse phase although it is much shorter than the progenitor phase that precedes it.
The detection of $\bar{\nu}_e$'s in the pre-SN phase is hence more suitable for the alert of the imminent supernova, which may be indeed possible a few days before core bounce for the Fe-core progenitors if neutrinos obey the normal mass hierarchy.

The electron-type antineutrinos from the 12 and 15 $M_\odot$ progenitors can be detected by all detectors, especially on the planned detectors such as Hyper-Kamiokande and JUNO if the distance to them is $\lesssim~1~\mathrm{kpc}$. 
The 9 $M_\odot$ progenitor will be quite difficult to observe with $\bar{\nu}_e$'s even if it is as close to us as 200 pc, the distance to Betelgeuse.
We may hence conclude that we can distinguish the two types of progenitors by detection or non-detection of $\bar{\nu}_e$ prior to collapse.
With DUNE, on the other hand, we will be able to detect more than a thousand of $\nu_e$'s from all the progenitor models if the distance to the source is again 200 pc and neutrinos have the inverted mass hierarchy. 
The event numbers are reduced by a factor of $\sim$10 if they obey the normal mass hierarchy.
Such a large difference in the number of detections suggests a possibility to judge the neutrino mass hierarchy.
It is interesting to see that the ONe-core progenitor offers the best chance in this case.
This implies that irrespective of the type and mass of progenitor we may be able to confirm our current understanding of the physics in the collapse phase.
Note, however, that $\nu_e$'s are not useless in distinguishing the progenitor types. Although it will not be easy observationally, the fact that $\nu_e$ emissions from the ONe-core progenitor in the pre-bounce phase occur in much shorter periods than those from the Fe-core progenitors may be utilized to discriminate the former from the latter.


Our estimates admittedly include several uncertainties.
In the following we comment on them in turn.
In this paper, we began the hydrodynamical simulations of the collapse phase when the central density becomes $\log_{10}{\rho_c}/[\mathrm{g\ cm^{-3}}] = 10.3$ for the Fe-core progenitors, which is rather arbitrary. 
In fact, the cores are already unstable at this point and have started to collapse in the quasi-static evolutionary calculations, which means that we could have switched to the dynamical simulations a bit earlier. 
Indeed, if we switch at $\log_{10}{\rho}_c/[\mathrm{g\ cm^{-3}}] = 10$, the time it takes to reach core bounce is shortened by more than a second.
This is due to artificially accelerated collapse in the new calculation, which is in turn caused mainly by differences between the EOS used in the stellar-evolution calculation and that employed in the dynamical simulation. 
The EC rates are also different. 
Although the discrepancy of more than a second in the time up to core bounce may seem not small, the difference in the event numbers may not be so large, since most of the deviation occurs immediately after the onset of the simulation, when the density is still not very high.


The uncertainty in the EOS also affects the EC rates through the mass fractions of heavy nuclei in the NSE composition.
\cite{buyu13} compared the nuclear composition of three multi-species EOS's including ours that are recently constructed for supernova simulations.
According to their results, differences in the mass fractions of heavy nuclei increase with temperature and/or density and become as large as a factor of two at $T = 2$ MeV and $\log_{10}{\rho}/[\mathrm{g\ cm^{-3}}] = 11$.
The different treatments of the surface, bulk and shell  energies of heavy nuclei are the main cause for the discrepancies. 
In fact, the temperature dependence of the shell energies that is incorporated in \cite{furusawa17b} tends to smooth out the mass distribution around closed shell nuclei and may reduce the EC rate at early times in the collapse phase by $\sim$20\% \citep{furusawa17}. 
The shell quenching considered in \cite{raduta16} may also affect the nuclear weak rates during the collapse phase.


As explained in section 3, we have employed the nuclear weak interaction rates obtained by detailed calculations for individual heavy nuclei whenever they are available.
As the density and temperature increase in the collapse phase, however, there appear heavy nuclei that are not included in these tables.
We are then forced to use for these nuclei the approximate formula, eq.(\ref{apro1}), for EC and another table \citep{tachi} for $\beta^-$ decay.
Since the approximate formula is based on the data of nuclei around $\beta$  stable line, it may not be applicable to neutron-rich nuclei.
The rates in Tachibana's table, on the other hand, are not meant for supernova simulations originally and calculated for isolated nuclei under the terrestrial condition.
We have hence included the Fermi-blocking of electrons in the final state very crudely.
Moreover, the data in this table do not include the contribution from excited states.
When the central density exceeds $\log_{10}{\rho_c}/[\mathrm{g\ cm^{-3}}]\sim 11.4$, most of $\bar{\nu}_e$'s come from the $\beta^-$ decays of nuclei, the rates of which are derived from this table. We certainly need to improve them in the future.  
In this paper, we have not treated the neutrino emissions at $\log_{10}{\rho}/[\mathrm{g\ cm^{-3}}] \gtrsim 13$ on purpose because nuclei become more and more exotic with their mass and atomic numbers getting larger to produce so-called nuclear pastas before uniform nuclear matter is realized.
The compositions and weak reaction rates of these nuclei are highly uncertain at such high densities.
Moreover, the dynamical simulations handle them in a very crude way, ignoring a possible variety of pasta configurations and interpolating the reaction rates between a certain sub-nuclear density and the nuclear saturation density.
As mentioned earlier, one can crudely estimate the number of detections of $\nu_e$ during the period from the time, at which $\log_{10}{\rho_c}/[\mathrm{g\ cm^{-3}}]=13$, until core bounce by simply extrapolating the event rates obtained in Figs.~\ref{fig10}-\ref{fig12}. 
We have found then that $\sim$200 more $\nu_e$'s may be observed by DUNE for the inverted hierarchy. We certainly need improvements in the treatment of this phase, which will be a future work.


Although it is much beyond the scope of this paper to take into account in detail the background noise for each detector and discuss the detection possibility quantitatively, we touch the issue briefly, since the actual detectability depends on it crucially. 
If we adopt several hundreds of events/day as the typical noise level of Super-Kamiokande at present, $\bar{\nu}_e$'s may not be detected even from FeCCSNe located at 200 pc. 
However, the background will be reduced remarkably to 0.21 events per hour after Gadolinium is doped as designed \citep{beacom04}.
An accompanied reduction of the energy threshold may increase the number of events by a factor of $\gtrsim$ 10 as demonstrated by \cite{yoshida16}.
The background for KamLAND is already very low $\sim$1 event/day and will not be a problem.
In the case of Hyper-Kamiokande, on the other hand, the reduction of the energy threshold, if possible, will have a big impact on the event number as mentioned earlier.

In this paper, we have considered only two relatively light Fe-core progenitors.
It is certainly important, though, to study other more massive progenitors systematically.
It should be also emphasized that the expected event numbers for the present models may change by a factor of a few if one considers various uncertainties in the current stellar-evolution calculation.
As stated at the beginning, our ultimate goal is to extend the current investigation until the end of the cooling phase of proto neutron stars seamlessly and consistently. 
It is stressed again that most of the studies on the neutrino emissions from CCSNe and their detections at terrestrial detectors done so far have treated the post-core bounce phase and the subsequent phase of the proto neutron star cooling separately and very little attention has been paid to the phase preceding them. 
Now that we have a lot of CCSN simulations that are successful to obtain explosions, we believe that we should make a serious effort to draw light curves and spectral evolutions of neutrinos that span the entire period from the progenitor phase up to the formation of the normal neutron star. 
This paper is just the first step.

\acknowledgments{
We are grateful to Dr. Tachibana for providing us with the table of nuclear weak interaction rates and Dr. Beacom for his useful advice.
This work is partly supported by Grant-in-Aid for Scientific Research from the Ministry of Education, Culture, Sports, Science and Technology of Japan 
(Nos. 24244036, 24103006, 26104007, 26400220, 26400271), and HPCI Strategic Program of Japanese MEXT. 
H.N. and S.F. are supported by Japan Society for the Promotion of Science Postdoctoral Fellowships　for Research Abroad.
Some numerical calculations were carried out on PC cluster at Center for Computational Astrophysics, National Astronomical Observatory of Japan.
K.T. is supported by Overseas Research Fellowships of Japan Society for the Promotion of Science (JSPS).}

\end{document}